\documentclass[12pt]{elsarticle}

\usepackage[textsize=small, disable]{todonotes}

\newcommand{\ivan}[1]{\todo[inline]{\textbf{Ivan: }\\#1}}
\newcommand{\marco}[1]{\todo[color=green!30,inline]{\textbf{Marco: }\\#1}}

\newcommand{\stefano}[1]{\todo[color=violet!30,inline]{\textbf{Stefano: }\\#1}}

\usepackage{ifthen}
\usepackage{comment}
\usepackage{tabularx} % for tables with tablegenerator
\usepackage{hyperref} % href basically

\usepackage[inline]{enumitem} % inline item enum and font change
\usepackage[title,toc,titletoc,header]{appendix} %appendix
\usepackage{xspace} %for marco space

\usepackage{amssymb}
\usepackage{amsmath}

\usepackage{cryptocode} 
\usepackage{footmisc}

\usepackage[ruled,linesnumbered, vlined]{algorithm2e}
\SetKwComment{Comment}{/* }{ */}
\SetKwInput{Witness}{Witness}
\SetKwInput{Instance}{Instance}
\SetKwInput{Parameters}{Parameters}

\makeatletter
\newcommand\footnoteref[1]{\protected@xdef\@thefnmark{\ref{#1}}\@footnotemark}
\makeatother

\usepackage{mathrsfs}

\usepackage{subcaption}

\usepackage{amsthm}
\usepackage{mdframed}  % For the box environment

\usepackage{fullpage}
\usepackage{fancyhdr, amsmath,amssymb}

\usepackage{caption}

\usepackage{hyperref} % for urls

\usepackage{booktabs}   % For professional table styling
\usepackage{multirow}   % For merging rows

\usepackage{adjustbox}

\usepackage{listings}

\definecolor{codebg}{rgb}{0.95, 0.95, 0.95} % Light gray background
\definecolor{keywordcolor}{rgb}{0.0, 0.2, 0.6} % Dark blue for keywords
\definecolor{stringcolor}{rgb}{0.6, 0.0, 0.0} % Dark red for strings
\definecolor{commentcolor}{rgb}{0.3, 0.5, 0.3} % Green for comments

\lstdefinestyle{solidity}{
    backgroundcolor=\color{codebg},
    basicstyle=\ttfamily\footnotesize,
    keywordstyle=\color{keywordcolor}\bfseries,
    stringstyle=\color{stringcolor},
    commentstyle=\color{commentcolor}\itshape,
    numbers=left,
    numberstyle=\tiny\color{gray},
    stepnumber=1,
    numbersep=5pt,
    showspaces=false,
    showstringspaces=false,
    frame=single,
    rulecolor=\color{black},
    captionpos=b,
    breaklines=true
}

\newcommand{\snark}{SNARK\xspace}
\newcommand{\snarks}{SNARKs\xspace}

\definecolor{spcolor}{gray}{0.28}

\newcommand{\defeq}{\vcentcolon=}

    \newcommand{\rel}{\mathcal{R}}
    \newcommand{\stat}{x}
    \newcommand{\wit}{w}

    \newcommand{\query}{application-specific query\xspace}
    \newcommand{\queries}{application-specific queries\xspace}

    \newcommand{\st}{\mathsf{st}}
    \newcommand{\tx}{\mathsf{tx}}
    \newcommand{\sok}{FC:ChaBalCha22}
    \newcommand{\Q}{\mathsf{Q}}
    \newcommand{\funcmr}{f_{\mathsf{MR}}}
    \newcommand{\verify}{\mathsf{Vrfy}}

    \newcommand{\lc}{\mathsf{L}}
    \newcommand{\fn}{\mathsf{FN}}
    \newcommand{\server}{\mathsf{S}}

    \newcommand{\app}{\mathsf{a}}
    \newcommand{\roottx}{\mathbb{R}_T}
    \newcommand{\rootfntx}{\mathbb{R'}_T}

    \newcommand{\pisnark}{\tilde \pi}

\newtheorem{definition}{Definition}

\journal{Blockchain: Research and Applications}

\begin{document}
%%
%% The "title" command  
\begin{frontmatter}

%\title{The Evolution of a Decentralized App:\\ Copyright Management with Scalable Verification}
%\title{Enabling Light Verification in a DApp:\\ Another Episode of Copyright Management\tnoteref{t1}}
%\title{A Tale on How to Decentralize an App:\\ the Case of Copyright Management}
%\title{The Whole Story on How to Decentralize an App:\\ the Case of Copyright Management}
%\title{A Tale on Increasing the Decentralization of an App:\\ the Case of Copyright Management}
%\title{Enabling Light Verification in a DApp:\\ the Case of Copyright Management\tnoteref{t1}}
% \title{Decentralization Chronicles: \\ the Evolution in the Case of Copyright Management \tnoteref{t1}}
\title{Efficient Query Verification for Blockchain Superlight Clients Using SNARKs}

\author[inst2]{Stefano De Angelis}
\ead{sdeangelis@unisa.it}

\author[inst1]{Ivan Visconti}
\ead{visconti@diag.uniroma1.it}

\author[inst1]{Andrea Vitaletti}
\ead{vitaletti@diag.uniroma1.it}

\author[inst1]{Marco Zecchini\corref{c1}}
\ead{zecchini@diag.uniroma1.it}

\affiliation[inst1]{organization={Sapienza University of Rome},
                    addressline={Via Ariosto, 25},
                    city={Rome},
                    postcode={00185}, 
                    country={Italy}}

\affiliation[inst2]{organization={University of Salerno},
                                  addressline={Via Giovanni Paolo II, 132},
                                  city={Fisciano},
                                  postcode={84084}, 
                                  country={Italy}}

%% Footnotes
\cortext[c1]{Corresponding author.}

%%
%% The abstract is a short summary of the work to be presented in the
%% article.
\begin{abstract}
      Blockchains are among the most powerful technologies to realize decentralized information systems. In order to safely enjoy all guarantees provided by a blockchain, one should maintain a full node, therefore maintaining an updated local copy of the ledger. This allows one to locally verify transactions, states of smart contracts, and to compute any information over them.

Unfortunately, for obvious practical reasons, a very large part of blockchain-based information systems consists of users relying on clients that access data stored in blockchains only through servers, without verifying what is received. In notable use cases, the user has application-specific queries that can be answered only by very few servers, sometimes all belonging to the same organization. 
This clearly re-introduces a single point of failure.

In this work we present an architecture allowing superlight clients (i.e., clients that do not want to download  the involved transactions) to outsource the computation of a query to a (possibly untrusted) server, receiving a trustworthy answer. Our architecture relies on the power of SNARKs and makes them lighter to compute  by using data obtained from full nodes and blockchain explorers, possibly leveraging the existence of smart contracts. 

The viability of our architecture is confirmed by an experimental evaluation on concrete scenarios. Our work paves the road towards blockchain-based information systems that remain decentralized and reliable even when users rely on common superlight clients (e.g., smartphones).   
\end{abstract}

% %%Research highlights
% \begin{highlights}
% \item Research highlight 1
% \item Research highlight 2
% \end{highlights}

\begin{keyword}
%% keywords here, in the form: keyword \sep keyword
Superlight Client \sep Blockchain Scalability \sep SNARK
%% PACS codes here, in the form: \PACS code \sep code
% \PACS 0000 \sep 1111
% %% MSC codes here, in the form: \MSC code \sep code
% %% or \MSC[2008] code \sep code (2000 is the default)
% \MSC 0000 \sep 1111
\end{keyword}

\end{frontmatter}

\ivan{un possibile titolo alternativo, che rilassa ``Efficient'' e si riferisce alla call for papers: 
Robust Blockchain-Based Information Systems with 
Superlight Clients}

\marco{uniformare superlight clients ovunque}

\section{Introduction}
\label{sec:introduction}

%\ivan{Let's be careful, we have cheated in this introduction mentioning storage issues for the super-light client, but since in the end we go for map-reduce queries this is unfair; indeed the super-light client can just download and store a single block, process it, and then continue with the next one, therefore there is no storage issue, just communication. Another pass on the intro is needed to make sure that we are not unfair}
%\marco{ho aggiunto un riferimento a questo in our contribution per essere più fair quanto introduciamo le map reduce queries}
Most of the emphasis in blockchain research has focused on designing and studying protocols capable of efficiently allowing a set of distrusting peers to achieve consensus on the status of the system, balancing efficiency (in particular, scalability) and security (in particular, increasing the number of participants in the consensus protocol) without relying on a third trusted party (TTP). The goal of the consensus protocol is to assemble transactions into blocks and order such blocks on the blockchain (ensuring a total order of transactions). Each transaction updates the state of the system according to some rules that can be either known a priori to all the participants (e.g., denying a user the ability to double-spend some cryptocurrency) or can be specified by users deploying smart contracts that can also be seen as programs running on the blockchain. However, not all peers have enough resources to store the blockchain and, therefore, locally access/read the ledger.

\paragraph{The RPC model}  To widen the audience of application developers and participants, full nodes of the blockchain expose the Remote Procedure Call (RPC) servers in order to allow users with constrained resources to access the blockchain. For example, Metamask\footnote{\url{https://metamask.io}}, which is a famous application that allows users to ease the management of their cryptocurrency and the interaction with the blockchain, makes use of JSON-RPC to connect to a single full node and access the blockchain. 
In general, this design has opened access to many developers, but it has also introduced again a centralization issue: developers have to trust the reply of the RPC. Recently, a cryptocurrency scam has been discovered that operates by tampering with Ethereum node RPC to defraud unsuspecting victims\footnote{\url{https://cryptopotato.com/slowmist-exposes-scam-using-malicious-rpc-node-modifications/}}. 
In the past, the above issue was not effective since originally, RPCs were designed to facilitate the access of off-chain applications developed locally on nodes participating in the consensus (hence, nodes that have autonomously checked the consistency and integrity of the blockchain). Instead, nowadays it is very common in many online application contexts (as shown by the Metamask case) to just rely on information received via RPCs without actually resorting to the chain of blocks to check the received answers.

In the above modern scenarios, users owning constrained devices do not have many more guarantees than in standard client/server applications. 

\paragraph{Application-specific queries}
Such centralization issues related to the RPC architecture could be obviously addressed by having constrained clients repeat the same queries (for instance, through RPCs) to multiple independent full nodes and compare their responses.

Of course, since these limited users are interrogating independent servers, this approach can work only on very general-purpose queries (e.g., state of a smart contract, existence of a transaction) that are standardized among different RPC servers. 
In some cases, servers can go beyond classic full node behavior and setup their own web services providing aggregated data which is of general interest for blockchain users (e.g., the number of transactions sent or received by an account or a smart contract), as in the case of blockchain explorers such as \href{https://etherscan.io/}{Etherscan} or  \href{https://www.blockchain.com/explorer}{Blockchain.com}. These web services, as well as RPC servers such as \href{https://www.infura.io/}{Infura}, have created a business activity on this information and, thus, they have all the interests in answering honestly to such queries, otherwise they risk losing customers (indeed, a user can interrogate different services comparing their answers and eventually detect bad behaviors).

Nevertheless, there is a large number of blockchain applications 
that might have their own specific interest in processing data, either because they want to retrieve filtered transactions with ad-hoc filters (e.g., a  user asks for all Bitcoin transactions with transferred value $>$1 BTC), or because they want the result of some ad-hoc aggregation (i.e., the average tip left by a transaction in 2024). 
A full node, with suitable incentives, could analyze the content of the transactions of the blockchain to answer such arbitrary queries. However, it is completely unclear how application developers requiring such sophisticated features can be sure to  successfully convince them to implement the desired additional logic.

For these reasons, in the wild\footnote{See for example, the queries elaborated through \href{https://dune.com/}{Dune Analytics} available at: \url{https://dune.com/queries/4069681} and \url{https://dune.com/hildobby/btc-etfs}},
% \marco{un caso d'uso credibile della blockchain, a parte DEFI (per cui servizi che fanno statistiche ce ne sono abbastanza), è come piattaforma di e-voting. Se durante il voto si raccolgono anche dei dati anagrafici, un servizio potrebbe essere mostrare statistiche sui dati anagrafici} 
% \ivan{sarebbe utile indicare qualche servizio in già in funzione per il quale si accede a server esterni; va bene DEFI, tracciability, evoting, qualunque cosa, ma bisognerebbe indicare il nome del servizio e dare un hyperlink}
there are natural useful queries that are neither handled through standard RPCs nor handled by well-known general-purpose services like blockchain explorers.
We refer to this type of queries as \emph{\queries}.     

Application-specific queries include requests %either (1) demand to retrieve many transactions matching a condition across blocks or (2) 
to which a full node cannot reply without implementing new functionalities. 
Individual servers or small groups (that are usually the organization in charge of a specific blockchain application) can then setup their own servers, that could be also full nodes, but more importantly are specifically designed to answer a specific set of \queries that are of interest for the application. However, since such servers are limited in number and could all belong to the same organization, this again introduces an even more urgent centralization problem for those users who are not able to setup their own full nodes.

\paragraph{The light client model} In order to 
avoid completely relying on full nodes, developers can set up light clients. Designed by Nakamoto's seminal paper~\cite{bitcoin}, a light client stores locally only a portion of the blocks of the blockchain (i.e., the headers) that is sufficient to guarantee the integrity of the rest of the block and to verify the correctness of the consensus protocol. In particular, in the headers, there is the root hash of a Merkle Tree built using as leaves the transactions of the block. In this way, a light-node client can ask a full-node one or more transactions (which are included in the part of the block that a light-node does not store) along with the Merkle proof used to attest that such transactions are actually in the ledger. Hence, light clients drastically reduce the amount of memory required for developers to verify the integrity of blockchain transactions or smart contract state. However, despite avoiding downloading and storing the entire blockchain, there are several important shortcomings.  

First of all, the amount of memory required by light clients still grows linearly with the size of the blockchain. Hence, eventually, the size of the chain of headers exceeds the memory available on constrained devices (e.g., smartphones). To work around this problem, Alchemy, a relevant company building blockchain software products, suggests pruning the chains of headers requiring only 400 MB.\footnote{\url{https://www.alchemy.com/overviews/light-node}} However, if a user wants to read some information that is stored in the pruned part of the ledger, she has to ask for such information from a full node without being able to verify it. 

Next, in case a light client is always online updating the chain of headers, there is a significant consumption of resources (e.g., the battery of a smartphone). If, instead, is only sometimes online, there are significant resources to use to get and check the missing blocks. 

Last, but not least, storing headers does not allow one to quickly verify the answer to a \query that affects a large number of transactions, since the light client should first download all the relevant transactions, then verify them, and finally run the query over them. Again, this can be excessive for a resource-constrained device (e.g., a smartphone with limited storage and battery charge).
% \ivan{I think it is worthy to mention the issue that storing headers does not allow to quickly verify the answer to a query that affects many transactions, since the client should first download all transactions, then verify them, and finally run the query over them.}

\paragraph{The goal of related work: verifying transactions saving on storage}
The need for a design that is more friendly to constrained devices is well known and indeed there is a relevant line of research \cite{SP:BKLZ20,FC:KiaMilZin20,AC:AFGK22,FC:KatBon23,ESORICS:LuTanWan20,FCW:KiaLamSto16,CCS:XZCZZJBS22, FC:VGSGJKOOT22, AFT:ANTZ23, FCTEMP:TZYT22} that tries to build light clients storing information of size \emph{sublinear} to the number of blocks of the blockchain. The works~\cite{FCW:KiaLamSto16, FC:KiaMilZin20, SP:BKLZ20} introduce, only for Proof-of-Work (PoW) based blockchain, the notion of Proofs of Proof of Work, which is a cryptographic proof ensuring in logarithmic complexity that a full node knows a chain containing a sufficient amount of work and provides different constructions of such a proof.  In~\cite{FC:KatBon23}, the authors propose, during the PoW consensus protocol, to reuse the solutions computed by miners that do not solve the puzzle of PoW in a specific round to generate a cryptographic proof guaranteeing the correctness of the new states of the blockchain. In \cite{FC:VGSGJKOOT22}, the authors leverage cryptographic proofs (of a different type with respect to Proofs of Proof of Work) to create periodic checkpoints on the state of the blockchain. This intuition is extended in~\cite{AC:AFGK22, CCS:XZCZZJBS22} where the authors devise proof systems enabling arbitrary updates for light clients. The works~\cite{ESORICS:LuTanWan20, AFT:ANTZ23,FCTEMP:TZYT22} describe interactive protocols running in a logarithmic number of rounds between the light clients and the full nodes to enable the former to penalize the latter in case of malicious behavior using the blockchain as an arbiter.
% \marco{rafforzare un po' qui per evitare di dover fare un paragrafo esplicito su related work}
More details about the state of the art on light clients can be found in a recent systematization of knowledge (SoK) provided by~\cite{FC:ChaBalCha22}.

\paragraph{Limits of prior work on light clients}
In general, as also analyzed in~\cite{FC:ChaBalCha22}, the above prior work faces the following question (see Section 3.1~\cite{FC:ChaBalCha22}): {\em is a specific transaction or account state included in the blockchain?}
However, there are cases in which the state of an application may depend on many transactions and, therefore, users must securely download them and, then, compute a function over them. This workload might not be affordable by a resource-constrained device.

\paragraph{Smart contract to store \query answers}
Smart contracts allow some application-oriented storage on the blockchain, in addition to the raw transactions. Full nodes and blockchain explorers allow fast access to such information. Therefore, one might think of keeping updated on the smart contract's state and the output of \queries. In this way, users running a light client simply need to access and verify one single state of the smart contract. 
%Indeed, light clients can be used to integrate the logic of applications with smart contracts one. 
%Indeed, light clients rely on the consensus to anchor data on-chain while efficiently responding to queries off-chain, addressing the specific needs of application-driven scenarios.

However, such an approach, while theoretically feasible, suffers from very severe practical limitations. First of all, storing and keeping updated a large amount of data (e.g., the results of a huge amount of interesting queries) on a smart contract impacts the cost that users have to support in order to interact with a blockchain application. Indeed, transaction fees are usually computed in relation to the effort (both in terms of storage and computational complexity) that validators have to do in order to execute such transactions. Second, it is reasonable to assume that, in some application contexts, interesting readings of the blockchain for users might arise dynamically while they are using the application (new features are added to web services very often). Unfortunately, many blockchain technologies (e.g., Ethereum Virtual Machine-based blockchains) do not natively allow one to modify the code of a smart contract once it is deployed, making it hard to store the answers to the new \queries \stefano{there are many techniques to update the execution of a smart contract or even change parts of their memory, e.g., Proxies or `surgeries` \url{https://github.com/BlossomLabs/surgery-app}}. Finally, some blockchain technologies, such as Bitcoin, do not have smart contracts and thus there is no help from the above techniques. 

\paragraph{A server answering \queries with cryptographic proofs}
A simple and widely used cryptographic proof consists of showing that a transaction exists in a block by providing the Merkle path from the root to the leaf. Considering a large number of transactions, this approach would 
require a huge amount of communication, and thus more advanced cryptographic proofs are required. 

In~\cite{CCS:XZCZZJBS22}, the authors propose to build a light client that, given a trusted block (e.g., the genesis block) of a blockchain, verifies that any subsequent block belongs to the longest chain with a cryptographic proof. Specifically, they propose to use a Succinct Non-Interactive Argument of Knowledge (\snark) which is a cryptographic proof allowing a user, called prover, to show to another user, called verifier, that she knows some secret information (witness) verifying a relation with some public information (claim). A \snark is succinct that means that verifying such proof requires an amount of time and space that is smaller than verifying the relation scanning the entire witness (i.e., the verification of a \snark has time and space complexity sublinear to the size of the witness). In~\cite{CCS:XZCZZJBS22}, the authors leverage this feature allowing a light client to verify that, given a trusted block, a subsequent block belongs to the longest chain with an amount of space and time that is sublinear to %the amount of data the light client would need to verify the same relation scanning the 
entire chain of blocks. % Note that, a \snark does not reveal any information about 

In theory, the above approach can be extended to certify  the answer for an \query and that all the information involved is taken from the blockchain. This means that the constrained client has a trusted block and a block updating the state of the blockchain % \ivan{not clear what the updating block is}
and the server creates a proof that guarantees that (a) the chaining of the entire blockchain is correct from the genesis to the last block, (b) the transactions needed to answer an \query are in the blockchain, and (c) the results of the query announced by the server are correct. Unfortunately, this solution has a very expensive cost to the server because computing \snarks for large blockchain updates can require extremely demanding computing resources. Indeed, as estimated by~\cite{FCTEMP:TZYT22}, maintaining such a system would cost approximately 50 million dollars per year, making it impractical for most use cases.

\paragraph{Open problem}

\begin{figure}
    \centering
    \includegraphics[width=0.7\linewidth]{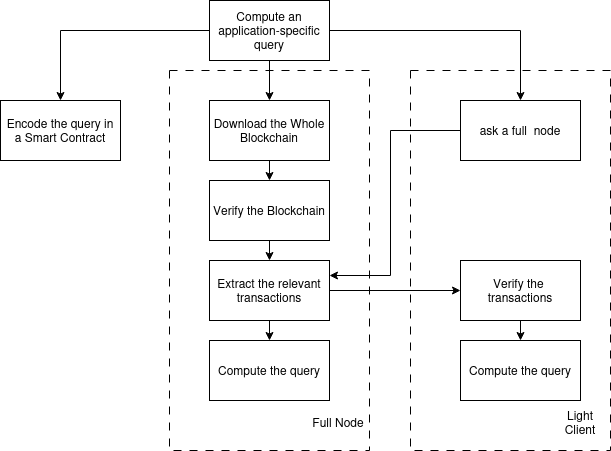}
    \caption{Current approaches to compute an application-specific query might be  unfeasible on resource constrained devices, such as smart phones. Implementing complex queries on smart contracts can be unfeasible or result in prohibitive costs. Smart phones do not have sufficient space to download the whole blockchain and when handling the query requires large bandwidth and/or memory, also light clients might be unfeasible.}
    \label{fig:current_approaches}
\end{figure}

Figure~\ref{fig:current_approaches} summarizes the current approaches for computing an application-specific query. Those approaches are not feasible on devices with limited resources, such as smartphones.

A practical approach to support such queries on smartphones is to rely on the collaboration of more powerful servers capable of implementing the queries, but this implies trusting those servers and clashes with the decentralization philosophy of the blockchain. To deal with this issue, smartphones might interact with multiple servers and rely on a majority argument, under the assumption that those servers have an interest in preserving their reputation by providing a good quality (i.e. trustable) service. 

%Furthermore, this approach requires that the group of servers providing the answers to the query are incentivized 
%\ivan{but this is true, there server of an application is incentivezed by the need of having the application on the blockchain; not sure this sentence adds anything }
%to deliver such an \emph{heavy} service. 

%\ivan{ho cambiato un po' il testo da qui fino alla fine della section, rileggere con attenzione}
However, when a new service starts, it is unlikely that
multiple distrustful servers are already interested in supporting it answering to complex queries.
Typically there is a bootstrap phase where the service is  managed by few servers all belonging to the same organization (or group of aligned organizations).
Then, once the service to compute the application-specific query has gained popularity, one can easily expect an increased interest of additional distrustful servers in answering those application-specific queries.
%is clear, but reaching a significant traction normally requires time; there is usually a ``bootstrap'' period where very few (it might be one) committed servers envision the future advantages and commit to deliver the service. 

During the above bootstrap phase, smartphones are supposed to trust a single server that might not even necessarily have an established reputation.  %In some contexts, it might be acceptable to trust a well-reputed full node, but  while we can expect that a full node provides access to the RPC, as already observed, the provisioning of other services should be motivated by some interest. \ivan{I think we discarded already full nodes, if not, we should it earlier anyway}
%\marco{mi sembra un po' brusco questo passaggio fra questi due paragrafi}

In light of the above discussion, blockchains currently fail to realize a general-purpose decentralized platform to answer \queries made by superlight clients (e.g., smartphones) that today represent the most widespread devices among the population.
%In light of the above discussion, the state of the art therefore indicates that blockchains fail to realize general-purpose decentralized platforms that can be accessed by lightweight clients. 
In fact, the only way to provide information to superlight clients consists of having, at least during a bootstrap phase, some trusted server (similarly to what clients currently do via RPCs) that provides application-specific answers to \queries. This is extremely unsatisfying, especially because the use of a blockchain is usually due to a desire for decentralization and resilience from single points of failure.  The above state of affairs motivates the natural open question of designing a decentralized architecture that allows superlight clients to efficiently and reliably obtain answers to \queries without relying on individual TTPs. However, it can still be acceptable to exploit independent full nodes or any other standard components of the blockchain architecture, that anyway are currently queried by a multitude of applications by standard RPC calls. 

%The open problem that we address in this paper consists specifically in showing an architecture allowing superlight client to participate in the bootstrap phase without trusting the server.

\paragraph{Our contribution} 
%Hence, during bootstrap, smartphones should trust a single server that might not even necessarily have an established reputation.
\stefano{Do we want to mention here that we ran an experimental evaluation with Plonky2?}In this work, we propose an architecture where superlight clients do not trust the server's answer as in a leap of faith, but instead get convinced by proofs generated by the server.

More in detail, our approach is based on the following main idea. The light client can verify the answer provided by the server through various verification steps that include the verification of succinct proofs, called \snarks, that are very fast to verify. While it is well known that the computation of  \snarks might be unfeasible on large datasets given in input to cryptographic hash functions, we present an effective solution allowing us to handle map-reduce \queries, employing a divide-et-impera approach capable of dealing with 
large sets of transactions that would be untractable in a single \snark. 
In particular, in our architecture, the server computes a \snark only for the blocks actually containing relevant transactions.  However, there is a subtle issue. A malicious server might cheat simply by omitting some important blocks, therefore, discarding some valuable transactions from the computation of the output of the query. To guarantee that all necessary transactions have been considered, the \snark is given by the server along with a link to some publicly available information, obtainable by standard components of the blockchain architecture, so that by inspection the superlight client is guaranteed that the only way to answer a query according to the publicly available information is to consider all relevant transactions. 
The above publicly available information is crucial for our architecture and is clearly application-specific. Therefore, we will focus on those scenarios where either the state of smart contracts, or RPCs, or blockchain explorers provide this public value. 
Note that since the nature of the above publicly available information makes it accessible through multiple nodes and therefore the superlight client can contact some of them and proceed in the computation if and only if all the replies are consistent.

To further improve the efficiency of our solution, the per-block \snarks are eventually merged into a single \snark that is verified by the superlight client. 

We stress that, while the map-reduce approach allows us to efficiently deal with \queries, it is not efficiently applicable to all possible kinds of queries.  This happens in particular when the computation of the query can not be split in subcomputations affecting only a subset of the relevant transactions.
While this means that our architecture does not provide a general-purpose solution, we make significant progress enabling several natural and significant use cases.

We first consider a use case where the \query is the computation of the average BTC transferred by a Bitcoin account, specifically the  %The super light client obtains the answer to such a query by a server that generates a proof ensuring the correctness of the queried result. The client, before verifying this proof, queries multiple blockchain explorers and full nodes (via RPC) to see how many transactions are sent or received by a specific address and to retrieve the block hashes of the blocks containing these transactions. Finally, once she gets this information, she verifies the proof for the retrieved number of transactions and for the retrieved block hashes. 
Satoshi Nakamoto’s account on Bitcoin that, at the time of writing, is involved in around 40k transactions. The experimental results show that our approach effectively verifies and computes the query using approximately $15$ MB. Answering the same query with a traditional Nakamoto light client requires around $700$ MB, while computing it with a light client that uses a sublinear amount of space in storage to keep track of the blocks (which is the goal of the mentioned track of research on light clients) requires around $200$ MB.
%\stefano{This is not what we implemented in the circuit though. We have an average calculation}

The second use case involves an on-chain election handled with a smart contract. The \query is about the average number  of votes for a specific candidate in a given time-interval. 
Our experiments, considering an election period spanning 7000 blocks (around one day) and 70000 votes, confirm that the approach effectively verifies and computes the query using approximately $9$ MB of data. Answering the same query with a traditional Nakamoto light client requires around $1.1$ GB, while computing it with a light client that uses a sublinear amount of space in storage to keep track of the blocks also requires around $1$ GB.

\begin{figure}
    \centering
    \includegraphics[width=0.8\linewidth]{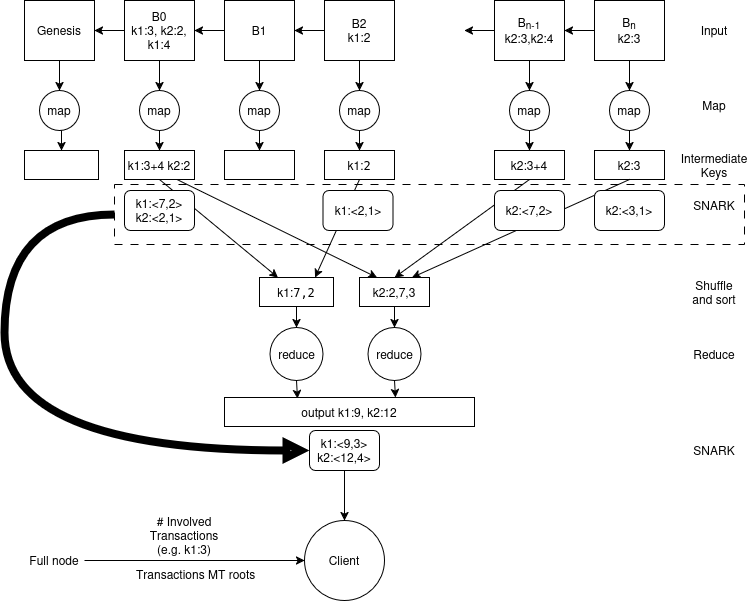}
    \caption{In this example we are interested in computing the average value of all the transactions generated by  users k1 and k2. For the sake of simplicity, we consider a mapper for each block. The SNARKs are used to prove the result of the computation. To prove all relevant transactions have been considered, the SNARK output is linked to a cumulative value available  in a full node; in the example the toal number of involved transactions. }
    \label{fig:MR}
\end{figure}

\paragraph{A concrete example}
Introduced in 2008 by~\cite{mapreduce}, Map-Reduce is a programming model for processing large datasets, in our case, the set of transactions in the blockchain. Figure~\ref{fig:MR} summarizes the main tasks of the model: 1) the Map task takes a set of data as input, in our case blocks and transactions, and converts it into a new set of data, where individual elements are broken down into intermediate key-value pairs; 2) the Reduce task takes the output from the Map as input and combines those intermediate key-value pairs into a smaller set of tuples used to eventually compute the output. 

% \ivan{for both the above use cases we specify what happens with our architecture, but we did not specify what would have happened using the available solutions. }
\section{Model}
\label{sec:model}

% \stefano{aggiunto da marco: in questa sezione, formalizziamo un po' più chiaramente quali sono i requirements e il threat model a cui risponde la nostra soluzione e che non è soddisfatta dallo stato dell'arte. mostriamo più nel dettaglio, quindi quali dei requirements non è soddisfatto dallo stato dell'arte}

\subsection{Preliminaries}
\label{subsec:preliminaries}
% In this subsection, we provide informal and intuitive definitions of a blockchain and of other cryptographic primitives that we are going to use in our solution. \stefano{can we remove this?}

\paragraph{\textbf{Notation}} For an array (or list) $X$ of size $n$, where $n \in \mathbb{N}$, we denote with $\{x_0, x_1, \dots, x_n\}$ the expanded list of its elements and with $|X|$ the size of the array. % The $i\text{-th}$ element is identified with $X[i] = x_i$, starting from 0. We represent the subsequence of indices from $i$ (included) to $j$ (excluded) with the notation $X[i\text{:}j]$; the notation $X[i\text{:}]$ represents elements from $i$ (included) till the end of the array, while the notation $X[\text{:}j]$ represents elements from the first index of the array till index $j$ (excluded). Negative indices represent elements from the end of the array, where $X[-1]$ is the last element.

For probabilistic polynomial-time algorithms $\mathsf{A}$ and $\mathsf{B}$, we denote by $\mathsf{P}\langle\mathsf{A}(a),\mathsf{B}(b)\rangle \rightarrow x$ the random process $\mathsf{P}$ obtained by having $\mathsf{A}$ and $\mathsf{B}$ interact on (private) auxiliary inputs $a$ and $b$, respectively, and with independent random coin tosses for $\mathsf{A}$ and $\mathsf{B}$. The value $x$ represents the output of $\mathsf{A}$ after the interaction. We extend this notation denoting with $\mathsf{P}\langle\mathsf{A}(a),\mathsf{B}_{1,\ldots,n}(b)\rangle \rightarrow x$, the random process $\mathsf{P}$ obtained by having $\mathsf{A}$ and $n$ executions of $\mathsf{B}$ interact on (private) auxiliary inputs $a$ and $b$ respectively, and with independent random coin tosses for $\mathsf{A}$ and every $\mathsf{B}_i$. The value $x$ represents the output of $\mathsf{A}$ after the interaction.

\paragraph{\textbf{Merkle Tree}} \stefano{as a side note: Ethereum does not use binary trees but Merkle Patricia Trees. Do you guys think it is relevant?}
A Merkle Tree computed over input values $x_1,\cdots,x_n$ is  a binary tree in which 
%collision-resistant (CR) hashes of 
the input values are placed at the leaves all with the same largest depth, and the value at each internal node is the collision-resistant hash of the values of its two children (or just the hash of the left child if the right child is missing). If $n$ is not a power of $2$, the right part of the tree will have missing nodes.
The height of the tree is logarithmic in the number of leaves.
The root of the Merkle Tree is a succinct representation of the entire sequence $x_1,\ldots,x_n$. We denote the root of a Merkle tree as $\mathbb{R}$.

% This model is well suited for parallel computations in distributed environments, where the map and reduce tasks can run on different nodes of the network. However, in this paper, we are more interested in using this model as a paradigmatic way to identify the kind of application-specific queries that we consider, namely queries that can be conveniently answered implementing the map-reduce model, irrespectively of the fact that the underlying tasks are distributed or not.

\paragraph{\textbf{Blockchain}}
We consider a blockchain network where validator nodes collectively maintain and update a valid chain $B$. A chain is \textit{valid} whether the honest majority of blockchain validators agreed on its content according to the network consensus rules. We also refer to a valid chain as \textit{canonical}. The chain maintains a list of transactions $T$ executed by the users uniquely identified with an account address, denoted with $\mathsf{acc}$. % , and a state $\st$ that encodes the view of $B$ for the users. % including the account balances and storage value of all users in the system.

A chain $B$ is an array of blocks, where the block $b_0$ is the genesis block. Every block includes a list of transactions. We denote the list of transactions in a chain as $T$, and with $T_{b_i}$ the transactions of block $b_i$. A transaction $\tx$ executes chain-specific tasks, e.g., transferring an amount of tokens from one account to another or invoking the execution of a smart contract. % \stefano{We might nor need the concept of state in this work.}

Every new block added to the chain updates the state of the ledger from $\st$ to $\st'$. Note that, for simplicity, the state of $B$ reflects the view of $B$ for the users, including not only the states of each account or smart contract that has sent or received a transaction but also other information available on the ledger, such as the block hash, the difficulty of a round of consensus, and the block number. 

The list of the transactions and the list of the states of each account or smart contract that has sent or received a transaction are leaves of a Merkle Tree. The block header maintains the transactions root. For a block $b_i$ we denote with $\mathbb{R}^{b_i}_T$ the transactions root. %, and $\mathbb{R}^{b_i}_\st$ the state root. 
We denote the inclusion of a block $b_i$ in a blockchain $B$ with $b_i \in B$. A transaction $\tx$ is said to be part of the canonical chain in the block $b_i$, if $\tx$ is a leaf of the Merkle Tree with root $\mathbb{R}^{b_i}_T$ and $b_i \in B$, and we denote it with $\tx \in B$.

\paragraph{\textbf{SNARKs}}
A proof system involves two parties: a prover $\mathcal{P}$ and a verifier $\mathcal{V}$. The prover processes a public claim $x$ along with a private witness $w$ satisfying a polynomial relation $\rel$ between $x$ and $w$ (i.e., $\rel(\stat, \wit) = 1$) and generates a proof $\pi$. The verifier then checks $\pi$ against $x$, and outputs $1$ if the proof is valid or $0$ otherwise.

A \emph{Succinct Non-Interactive Argument of Knowledge} (\snark) is a specific type of proof system providing these key properties:
\begin{itemize}
    \item \emph{Non-interactivity:} $\mathcal{P}$ sends only one message containing $\pi$ to $\mathcal{V}$ with no further communication required.
    \item \emph{Completeness:} if $\rel(\stat, \wit) = 1$ and $\mathcal{P}$ honestly generates $\pi$, then $\mathcal{V}$ verifying $\pi$ outputs $1$ with overwhelming probability.
    \item \emph{Knowledge Soundness:} if $\rel(\stat, \wit) = 1$ and $\mathcal{V}$ verifying $\pi$ outputs $1$, then $\mathcal{P}$ knows $\wit$ for $\stat$ with overwhelming probability.
  \item \emph{Succinctness:} The size of the proof $\pi$ is sublinear in the size of $\wit$ and to verify it requires time complexity that is linear to size of $\stat$ plus a value that is sublinear in the size of $\wit$. 
\end{itemize}

Prover and verifier run their respective algorithms having as input a set of parameters, used as a common reference string. Depending on the different \snark constructions, such parameters can be generated autonomously by the parties or must be generated by a third party that generates the common reference string with some secret that, then, is trusted to destroy. 

A \emph{recursive} \snark system leverages these properties so that one proof can verify the correctness of other proofs within its own proving circuit. In this paradigm, a single proof $\pi$ attests that several proofs $\pi_1,\ldots, \pi_n$, each with a corresponding public claim $x_1,\ldots, x_n$, have been validated correctly. % Intuitively, the prover constructs
% \[
% \mathcal{P}(x_1,\pi_1,\; x_2,\pi_2,\;\ldots)\;\;\rightarrow\;\;\pi,
% \]
When the outer proof $\pi$ is verified, it implies that all proofs $\pi_1,\ldots, \pi_n$ are also valid. %, i.e.,
% \[
% \mathcal{V}(x,\pi) \rightarrow 1 \; \mbox{\textit{if and only if }}  
% \mathcal{V}(x_1,\pi_1) \rightarrow 1,\; 
% \mathcal{V}(x_2,\pi_2) \rightarrow 1,\;\ldots
% \]

\subsection{Map-Reduce Blockchain Queries}
% \ivan{descrivere cosa sono in modo da stabilire anche la notazione che usiamo poi quando descriviamo il nostro sistema}
Introduced in 2008 by~\cite{mapreduce}, Map-Reduce is a programming model processing large datasets that can be used in a broad variety of real-world tasks.
%\marco{darei intuizione generale di come funziona e poi come si usa nel nostro caso}

In the context of blockchain-based queries, the dataset is the canonical chain $B$ and each query accesses multiple blocks.

%Blockchain queries for map-reduce computation have been informally introduced in \cite{10.1145/3658644.3670354}. \marco{non riesco a trovare il riferimento a questa cosa nel paper... va spiegato perchè facciamo questa conclusione} 
These queries are decomposed into smaller, independent sub-queries that perform a generic computation of a function
over a sublist of transactions. For example, the average of tokens per transaction transferred from a blockchain account  can be computed as a map-reduce query that splits the calculation across relevant blocks, i.e. all those blocks that include relevant token transfer transactions.

In our work, we formalize the definition of map-reduce blockchain query. A map-reduce query is an operation made of three steps.
The first two steps follow the paradigm proposed by~\cite{mapreduce} (see Section 2):% , we express the computation of a map-reduce blockchain query with the two functions:

\begin{itemize}
    \item $\mathsf{map}$: takes as input a key/value pair and outputs a set of intermediate key/value pairs. Its main goal is to split a large computation in smaller tasks (e.g., from the entire blockchain to one computation per block) and produce a set of intermediate key/value pairs for each task (e.g., one set per block) computed according to a function $f_{\mathsf{M}}: \{0,1\}^* \rightarrow \{0,1\}^*$.
    \item $\mathsf{reduce}$: takes as input an intermediate key and a set of values for that key and outputs a smaller set of values obtained by merging and combining together the values of the input set computed according to a function $f_{\mathsf{R}}: \{0,1\}^* \rightarrow \{0,1\}^*$. Its main goal is to allow to compute a function over a set of values too large to fit in memory in a sequential way.
\end{itemize}

% Finally, in the third step, a map-reduce query uses an aggregation function that combines the partial results from the map-reduce computation and returns the final answer to the query. 

In the following, for the sake of simplicity, we define a map-reduce query only over the set of transactions in a blockchain $B$, however, the domain for this function can be easily extended also to other information of the ledger.
Formally, a map-reduce blockchain query is a function $\funcmr: B \rightarrow \{0,1\}^*$ defined over the canonical chain $B$. The returned value of this function is a binary string $\{0,1\}^*$. % list of \textit{key/value} pairs, where the \textit{key} is a block identifier $b_i$ and the \textit{value} is a result represented as a binary string $\{0,1\}^*$ computed over the block $b_i$.

\subsection{System Model and Properties}
\label{subsec:properties}
%\marco{Aspetto a ripassare su questa sezione la stabilizzazione di map reduce queries}
In this subsection, we introduce and formalize which parties are involved in the system and which status of the system they can reach (and cannot reach) by defining the properties of the system. 

In the system, there are three types of users:

\begin{description}

    \item[Full Node] A full node is a blockchain node that maintains the full history of the chain $B$ including the blocks, and their updated states. After a bootstrap period, every full node is able to remain synchronised with the longest valid chain. Therefore, the amount of data a full node downloads when joining the blockchain is linear in bandwidth and storage with the length of the chain $|B|$. Full nodes can autonomously query the chain, and execute transactions and smart contracts functions. These nodes compute a set of function queries $f_1, \ldots, f_m$ about the ledger and make the output available to superlight clients. Such functions can be computed either on raw data (e.g., retrieving a block header, the details of a transaction included in the chain) or on more sophisticated but of general interest data as in the case of blockchain explorer (e.g., balance of an account, token transfer histories, account-level analytics). Full nodes expose either an RPC interface over a communication protocol like HTTPS or WebSocket or a user-friendly web sites (as in the case of blockchain explorer) to provide to the superlight clients the output of $f_1, \ldots, f_m$. Each of these functions takes as input $B$ and outputs a state for the blockchain $\st$.
    
    \item[Oracle Server] An oracle server, referred to as an Oracle, % \stefano{oracles might operate a full node or use a third party service. They are not operated by a full node.} \marco{vedi se ti suona con extends the functionalities} 
    extends the functionalities of a full node and exposes map-reduce queries that are not available in the other full nodes because these are application-specific and require the execution of intense computing tasks. These servers leverage high-performance infrastructures with minimal constraints on bandwidth, storage, or computation. 
    Oracles maintain access to the entire history of the blockchain, either by querying full nodes or running their own instance.
    
    \item[Stateless Superlight Client (SSLC)]  A computationally constrained node (e.g., smartphone, laptop) that aims to track a specific state or answer a \query over an application or account on the blockchain, without downloading the full chain history of size $|B|$. We define this type of client also as \textit{stateless}, because it does not need to maintain the blockchain state locally once it has verified a \query.
    % \stefano{Alternatie description: A resource-constrained node (e.g., a smartphone or smartwatch) that aims to interact with a particular blockchain application without dowloading large amount of data, or maintaining the full state of the blockchain locally. Ideally, this light client queries the exeuction of a map-reduce query to an Oracle, and processes the result with minimal computation and storage overhead. We define this type of client \textit{stateless}, becasue it does not need to maintain the blockchain state locally, and \textit{super light}, because only process data that is sublinear to the chain $C$.} FATTO MERGE
\end{description}

In our model, we consider queries that are map-reduce ones and that are application-specific which means that are not supported by general-purpose full nodes.  In other words, the output of the query is not obtained by simply running $f_1, \ldots, f_m$. While full nodes allow retrieving transactions, they are not supposed to compute their content, especially when the computation is not a general-purpose one. Typical examples of queries that are not supported by full nodes consist of filtering the transactions based on their content and, finally, computing a function using as input the content of filtered transactions. Formally, given a blockchain $B$, we consider a map-reduce query $\funcmr$ such that $\funcmr \not = f_i, \forall i \in  \{f_1, \ldots, f_m\}$.
% \stefano{now that we defined map-reduce query we might not need this part}
% \marco{una application specific query è una query a cui non viene fornita una risposta da un full node (per altro, alcune rpc call possono essere map reduce).. ovviamnete è una parte che va levata ma non perderei questa nozione}

In the following, we present a model for our system that is inspired by~\cite{\sok}. Given a SSLC $\lc$, an oracle server $\server$ and a set of full nodes $\fn_1, \ldots, \fn_n$, our model presents the following protocols:
    \begin{description} 
        \item[$\mathsf{Q}\langle \lc(\funcmr), \server(B) \rangle \rightarrow (r, \pi)$] To compute the answer to an \query, a SSLC $\lc$, taking as input a function $\funcmr$ such that $\funcmr \not = f_i, \forall i \in  \{f_1, \ldots, f_m\}$, runs an interactive protocol $\mathsf{Q}$ together with an Oracle server $\server$, taking as input a blockchain $B$, and that outputs a reply $r \defeq \funcmr(B)$ and a proof $\pi$. % and the blockchain state $\st$. 
        % \ivan{non chiaro il ruolo esplicito di st, visto che poi infatti finisce in input insieme a $\pi$ alla prossima procedura; secondo me e' un elemento specifico della nostra costruzione ma in quanto tale non dovrebbe condizionare una definizione che invece è generica; st è dentro $\pi$}
        \item[$\mathsf{Vrfy}\langle\lc(\funcmr, r, \pi), \fn_{1, \ldots, n}(B)\rangle \rightarrow b$] To verify the correctness of an answer to an \query, a SSLC $\lc$, taking as input 
        % a blockchain state $\st$,
        a function $\funcmr$, a reply to an \query $r$ and a proof $\pi$, runs an interactive protocol $\mathsf{Vrfy}$ together with full nodes $\fn_1, \ldots, \fn_n$, taking as input a blockchain $B$, and outputs $b \in \{0, 1\}$ such that if the proof is correct $b = 1$; if the proof is not correct, $b = 0$. %; the value $\bot$ indicates that $\lc$ has rejected the execution.
        % \ivan{non chiara la necessita' di distinguere $\bot$ da $0$; di solito è importante che l'output sia 1 se tutto ok, 0 otherwise}
    \end{description}

Our model makes use of the following assumption:
    \begin{description}
        \item[Existential Honesty] There is at least one honest full node in the set of full nodes. Such assumption ensures that, given a blockchain $B$ with state $\st$, a SSLC contacting independent full nodes $\fn_1, \ldots, \fn_n$ during $\verify()$ gets the same $\st$; otherwise, if it only receives one $\st'$ such that $\st' \not = \st$, $\verify()$ outputs $0$ and $\lc$ restarts it with a different set of full nodes.
    \end{description}
    
    Our model ensures the following properties:
    \begin{description}
        \item[Secure Querying] Given a blockchain $B$ with state $\st$ that can be obtained by $\lc$ contacting $\fn_1, \ldots, \fn_n$, for any \query function $\funcmr$ involving the transactions $\tx_1, \dots, \tx_k$, it is computationally hard for a malicious oracle server to generate a proof $\pi$ for a reply $r'$ such that $\mathsf{Vrfy}\langle\lc(\funcmr, r', \pi),$ $ \fn_{1, \ldots, n}(B)\rangle = 1 \land (\exists i \in \{1, \ldots, k\} \mbox{ s.t. } \tx_i \not \in B \lor \funcmr(B) \not = r')$.

% \ivan{penso che dalla notazione in poi ci siano spesso incoerenze sul fatto che si parta da 0 oppure da 1 così come }
        
        \item[Efficiency] We identify the following properties in terms of storage, computation, and communication costs.
        \begin{itemize}
            \item \emph{Efficient storage:} The amount of memory needed by the SSLC to run $\Q()$ and $\verify()$ is sublinear to $|T|$.
% \ivan{il numero di transazioni in un blocco è una costante? se si noi non otteniamo efficient storage visto che per una query potremmo avere gli header di tutti i blocchi, o mi sbaglio?}
            
            \item \emph{Efficient communication:} $\Q()$ and $\verify()$ requires communication costs that is sublinear to $|T|$.
% \ivan{stesso commento di prima}            
            \item \emph{Efficient client computation:} $\Q()$ and $\verify()$ requires computation costs for the SSLC that are sublinear to $|T|$.
% \ivan{stesso commento di prima}
            
        \end{itemize}
    \end{description}

\paragraph{Reputation of full nodes} We have adapted to our case the Existential Honesty assumption from~\cite{AFT:ANTZ23}. Full nodes (such as Infura) or blockchain explorers (such as Etherscan) have founded business activities on their ability to maintain and answer queries to users with constrained devices and, thus, they have all the interests in answering honestly to such queries; otherwise, they risk losing customers (indeed, a user can interrogate different services comparing their answers and eventually detect bad behaviors). Hence, in our model, full nodes are interested in the reputation they have with the other users in the system. For this reason, it is unlikely that they answer dishonestly to queries made by a SSLC,
therefore the Existential Honesty assumption described above is concretly realistic.  

\paragraph{Differences and similarities with~\cite{\sok}} As previously mentioned, the model presented in this section is inspired by the definition of Section 3.1 of~\cite{\sok}. However, we deviate from their work to make the model more suitable for the context (i.e., we consider \queries) that we are considering. In particular, our model enables us to answer queries related to many transactions and a function $\funcmr$ maintaining the same efficiency properties. % \ivan{queste cose appena dette riguardano la progettazione del sistema non la descrizione del modello}
Our system relies on the Existential Honesty assumption which is an extension of one mentioned in~\cite{\sok} called ``Game-theoretic assumptions''. Differently from them, our model gives more detail about this assumption regarding the behavior of full nodes and how this affects our system. Given this assumption, the SSLCs in our model do not need to go through any bootstrap or sync phase (i.e., starting from a known genesis block, SSLCs update and verify the state $\st$ of $B$). Even though these phases enhance the security of the system, we believe that the assumptions of our model are reasonable in many real-world scenarios. However, our model can be easily modified to include a bootstrap or sync phase, as in~\cite{\sok}.

\section{Stateless Superlight Clients} \label{sec:solution}

% \stefano{From here, some notations might be disaligned with the model. We need to re-align them}
\stefano{Although I am the first supporter of "stateless", here we do have some kind of "state" downloaded by the light client. Do you guys think it can confuse the reader? (it confuses me a bit)}
\marco{perchè è stateless è spiegato nella sezione precendete ed è perchè fra due sessioni diverse non mantiene lo stato ma lo butta via... non so se ti riferivi a questo}
In this section, we show how to realize a \textit{Stateless Superlight Client (SSLC)}, a solution that enables light clients to execute application-specific queries on blockchain data with minimal trust, computation, and storage requirements. % Unlike light client designs~\cite{bitcoin, FC:KiaMilZin20,AC:AFGK22,FC:KatBon23,ESORICS:LuTanWan20,FCW:KiaLamSto16,CCS:XZCZZJBS22, FC:VGSGJKOOT22, AFT:ANTZ23, FCTEMP:TZYT22}, which maintain a persistent state (e.g., a valid chain of block headers, state commitments, Merkle roots), SSLCs operate without storing historical blockchain data after.
At their essence, an SSLC is a resource-constrained device running a blockchain-specific application that requires the execution of computationally expensive queries on a blockchain $B$. % The light client assumes the existence of a function $f(\cdot)$ that performs heavy computation over historical blockchain data. 
We restrict the scope to map-reduce blockchain queries $\funcmr$ for a canonical chain $B$. We idealize these functions to be available through an oracle that provides computational results on demand.
This assumption allows us to analyze the security and correctness of the protocol without requiring the verifier to perform the full computation, relying instead on succinct SNARK proofs of correctness.

\begin{figure}[h!] % Placement options: h (here), t (top), b (bottom), p (page)
    \centering
    \includegraphics[width=0.8\textwidth]{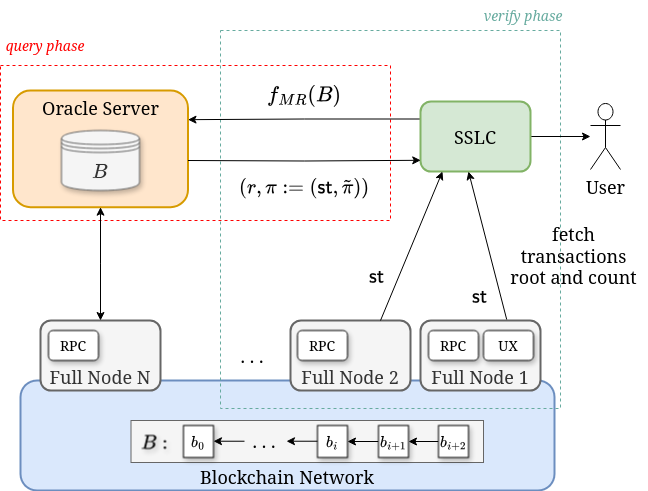}
    \caption{Stateless Superlight Client Architecture.}
    \label{fig:sslc-arch}
\end{figure}

%  \stefano{Secondo me definire ``stato" il transaction root + un counter arbitrario di transazioni (che possono essere 1 o 1000) è una over-semplificazione. Un lavoro abbastanza famoso dove viene definito lo stato della blockchain è questo \url{https://eprint.iacr.org/2017/1070.pdf}}

Figure~\ref{fig:sslc-arch} provides a graphical representation of the solution. The SSLC (verifier) engages in an interactive protocol with an application-oriented Oracle Server (prover) and general-purpose full nodes that maintain a valid blockchain $B$. The protocol consists of two phases: \textit{query} and \textit{verify}. In the query phase, the verifier requests the execution of a map-reduce query to the prover. The prover runs the computation and returns the result $r$ along with some information about the blockchain $\st$ and \snark proof $\pisnark$ that attests to its correctness.

In the verify phase, the verifier connects to different general-purpose full nodes to collect relevant information about the ledger $\st$ to verify the proof, such as the root of transaction Merkle trees in each block involved in the computation and the total number of transactions allegedly used. As we will see, this information is crucial for the security of the protocol; without it, the client cannot ensure that the prover has correctly accounted for all valid transactions in the canonical chain that were involved in the computation.
% \marco{stesso discorso per i due paragrafi precedenti su $\st$}

It is worth noting that the number of downloads of transaction roots is sublinear in the number of transactions processed to compute the final result $r$, assuming that each computed block includes a large number of transactions involved in the computation. 
% \ivan{forse possiamo sanare la questione di sublinear offrendo numeri concreti; mi spiego: se una blockchian ha n blocchi ed in ogni blocco ci sono mediamente m transazioni, la tablia della blockchain è n*m; se m è una costante non possiamo dire di essere sublinear; quindi dovremmo dire che m non è costante; per dire questo, dobbiamo far riferimento alle blockchain classicare e contare il numero di transazioni per blocco e verificare che è ad esempio enormemente maggiore log n, magari è SQRT(n); riusciamo a fornire qualche esempio per Ethereum e Bitcoin? questo sanerebbe la nostra deisderata sublinearità}
Therefore, with this approach, we ensure that the communication complexity remains proportional only to the number of connections to full nodes while the amount of downloaded data remains sublinear to the growing chain. We will see that for application-specific use cases, this approach is optimal, as it allows light clients to download a significantly smaller amount of data.

In the next section, we present in detail the SSLC solution. We first proceed in Section~\ref{sec:mr-comp} with the description of how to compute the map-reduce queries in our context. % and put some security requirements to ensure completeness. 
Then, in Section~\ref{sec:solo-proof}, we show how to compute the \snark proof attesting to the correctness of the computation of the function $\funcmr$.
Finally, in Section \ref{sec:sol-proto}, we provide an instantiation of the SSLC and a detailed description of the protocol. 

\subsection{Map-Reduce Query Computation} \label{sec:mr-comp}

The oracle server maintains a copy of the canonical chain $B$ and provides an interface for the SSLC to execute map-reduce blockchain queries. The computation of a map-reduce query is application-specific and for a blockchain application corresponds to the implementation of the function $\funcmr(B)$. Obviously, for an application, there might be more than one map-reduce query available. 
The function selects and operates only on specific blockchain transactions, such as those sent or received by one or more blockchain accounts or one or more smart contracts. 

We describe the computing steps of the map-reduce query in Figure \ref{fig:map_reduce_steps}. For the sake of simplicity, in Figure~\ref{fig:map_reduce_steps}, we show how to compute a map-reduce query for transactions involving a single blockchain account ($\mathsf{acc}_1$), but the described steps can be easily applied also to the case with more blockchain accounts or smart contracts.

\medskip

\begin{minipage}{\textwidth}
\begin{mdframed}[]
\centering \textbf{Algorithm to compute $\funcmr$} % Title of the protocol
\medskip

\begin{tabular}{@{}p{0.95\linewidth}@{}}  % Single column for alignment
\hline\hline

% The Oracle Server splits the computation into $i \in |B'|$ $\mathsf{map}$-$\mathsf{reduce}$ tasks. \\
We assume we are interested in computing a function $\funcmr$ for the transactions involving a specific account $\mathsf{acc}_1$: % and $\mathsf{acc}_2$:

\medskip

\textbf{Map Phase $\mathsf{map_i}$:} % $\mathsf{map_i}(b_i, T_{b_i}) \rightarrow \mathsf{MR_i}$  
it takes as input a block $b_i$ as key and the set of transactions of that block $T_{b_i}$ as value and
\begin{enumerate}
    \item In $T_{b_i}$, identify the set of transactions $\{\tx_1, \ldots, \tx_{k'}\}$ sent or received by $\mathsf{acc}_1$;
    \item Compute a function $f_\mathsf{M}$ on this set of transactions and save its value in a variable $v_i$;
    % \item Group the result by key using the block identifier $k = b_i$.
    %\item Repeat the previous two steps for $\mathsf{acc}_2$ and store the output of $f_\mathsf{M}$ in $v^i_2$
    \item Output the pair $(\mathsf{acc}_1, v_i)$ %$, (\mathsf{acc}_2, v^i_2)$.
\end{enumerate} \\
\textbf{Reduce Phase $\mathsf{reduce}$:} % $\mathsf{reduce}(b_i, \mathsf{MR_i}) \rightarrow (b_i, \mathsf{RR_i})$  
it takes as input a blockchain account $\mathsf{acc}_1$ as key and a set of values $\{v_1, \ldots, v_{|B|}\}$ computed during the map phase and 
\begin{enumerate}
    % \item Extract the values from $\mathsf{MR_i}$, $\mathsf{list}(v) \leftarrow \mathsf{MR_i}[b_i]$.
    \item Combines the values $\{v_1, \ldots, v_{|B|}\}$ into a single result $v$ by applying the function $f_{R}$.
    \item Output the pair $(\mathsf{acc}_1, v)$.
\end{enumerate}
\end{tabular}
\end{mdframed}
\end{minipage}
\captionof{figure}{Computing steps of a Map-Reduce query.}
\label{fig:map_reduce_steps}

\setlength{\parindent}{15pt} % Restore indentation after mdframed

\smallskip

\subsection{A \snark Proof System for the Oracle Server and the SSLC}
\label{sec:solo-proof}

% \paragraph{\textbf{Trust-minimized Oracle Server with SNARKs}} 

A naive approach to query verification requires the oracle to provide a list of all transactions used to compute the result. The light client could then independently query full nodes to verify their presence in the blockchain. However, this approach results in query complexity linear in the number of transactions, making it impractical for large-scale computations. For example, a simple P2PKH (Pay-to-PubKey-Hash) transaction on the Bitcoin network with one UTXO input and one output is large ~200 bytes \cite{DBLP:journals/corr/abs-2102-12796}, and the Bitcoin RPC call $\mathsf{getwartransaction}$~\cite{bitcoin_getrawtransaction} with the full verbose format will add ~1KB metadata, leading to 1.2KB per transaction. For a computation involving 1M transactions, the client will download ~1.2GB of data from each full node, which is unrealistic for a resource-constrained device.

Instead, we use SNARK proofs, which allow a prover to generate a succinct proof of correct computation. The oracle server computes the map-reduce query in sub-tasks and, for each sub-task, returns the result alongside a SNARK proof $\pisnark$. 
The proof must ensure that (i) the function $\funcmr(\cdot)$ was correctly executed, and (ii) each transaction used belongs to the transaction root $\mathbb{R}^{b_i}_T$ of some block $b_i$. 
By proving the existence of transactions within their respective blocks, the SSLC only needs to check the existence of those blocks in $B$. % and ensure the total transaction count matches the completeness criterion.

% \marco{prima introdurrei tutto poi proverei perchè garantisce la sicurezza del modello}
% This design achieves two main objectives. First, it ensures \textit{secure querying}, as the prover cannot forge a valid proof unless it has correctly performed the computation on a predefined verifiable set of blockchain transactions. Second, it preserves \textit{efficiency}, as the verifier only needs to download the SNARK proof and the total number of transactions involved in the computation to reach a trust-minimized conclusion. By replacing linear verification steps with a succinct proof system, our approach minimizes the resource consumption for an SSLC, making it suitable for resource-constrained devices that cannot afford to process large volumes of blockchain data.

\medskip

In the rest of this section, we describe the proof system of our SSLC interactive protocol. 
As we said above, the prover produces a SNARK per $\funcmr$ computation. % Practically, for every sub-task and block $b_i$, the prover runs an arithmetic circuit that:
% \stefano{I am a bit confused here. Why we need $\roottx^{b_{i_1}}, \ldots, \roottx^{b_{i_m}}$? Is the same root for block $b_i$ here? Otherwise, if we consider more block roots, isn't $\roottx^{b_1}, \ldots, \roottx^{b_m}$ sufficient? 
% \\
% Also, having $|B|$ does it mean that we are taking roots from all blocks of the canonical chain $B$? I thought we worked on a subset $B'$ with $|B'| < |B|$}
In particular, given a set of transaction roots $\{ \roottx^{b_{i_1}}, \ldots, \roottx^{b_{i_m}}\}$ with $1 \le i_1 < i_2 < \dots < i_m \le |B|$\footnote{\label{fn:root}This set of transaction roots is composed of transaction roots belonging to blocks that are not necessarily subsequent in the ledger.}, a public claim $x \defeq (\{ \roottx^{b_{i_1}}, \ldots, \roottx^{b_{i_m}}\}, r)$ and a private witness $w \defeq (B)$, an oracle server is able to generate a SNARK proof $\pisnark$ when $x$ and $w$ satisfy the following relation $\rel$:

\begin{itemize}
\item $\funcmr(B) = r$ and
\item each transaction $\tx$, used to compute $\funcmr$ and contained in a block $b_i$ of $|B|$, is a leaf of the transaction Merkle Tree identified with $\roottx^{b_i}$. %ensures \textit{transaction existence} by checking the Merkle authentication path of each transaction with the canonical root $\mathbb{R}^{b_i}_T$
% \item ensures that all processed transactions are unique and satisfy $\rho(\tx, \app) = 1$.
\end{itemize}

% Broadly, the prover of the SSLC protocol aims to convince the verifier that for each processed block $b_i$, the following claim is true:

% \begin{quote}
%     \textit{Claim (as a prover): I computed a sub-task of a map-reduce function $\funcmr$ on a set of $|T'_{b_i}|$ distinct transactions from the canonical block $b_i$ with transactions header $\mathbb{R}^{b_i}_T$ such that the result of the computation is $f_{MR_i}(T_{b_i}) = RR_i$.}
% \end{quote}

To compute such \snark proof and ensure secure querying in our protocol, we need to consider two issues. 
First of all, a dishonest oracle server may deviate from the correct computation of $\funcmr$ in various ways (e.g., she can discard some transactions during the computation or use multiple times a single transaction during the computation) still producing an accepting proof $\pisnark$.
In this section, we are presenting some techniques to ensure the secure querying in our protocol.

Second, $\funcmr$ could be computed over many transactions (in the order of magnitude of thousands or tens of thousands of transactions) and, therefore, computing such proof for such a large claim may require enormous computational resources for the oracle server or even be infeasible. In the following, we present a technique to improve the scalability for the prover exploiting the map-reduce paradigm and, in particular, its capacity of dividing the load of computing a larger task into smaller sub-tasks.

% \paragraph{\textbf{Transactions Uniqueness via Ordered Merkle Proofs}} 
\paragraph{\textbf{Preventing malicious behavior of the oracle server}} 
A dishonest server may deviate from the correct computation of $\funcmr$ while still being able to produce a proof $\pisnark$ that is accepted by an SSLC when:  
\begin{itemize}
    \item the server does not include some transactions during the computation of $\funcmr$ that should be included (because these involve the account $\mathsf{acc}_1$);
    \item the server includes multiple times the same transaction $\tx$ omitting others in the same block;
\end{itemize}

To ensure that the oracle server includes all the transactions to compute $\funcmr$, the SSLC must compare, during the $\mathsf{Vrfy}(\cdot)$ protocol, the $\st$ received by the oracle server with the values of $\st$ retrieved by a set of independent full nodes. This means that $\st$ has to contain information on the cardinality of transactions involved in the computation of $\funcmr$, such as the number of transactions, and that can be easily retrieved by full nodes (for example, with RPC calls). We will discuss more details about this aspect in the next section. 

To guarantee the uniqueness of processed transactions, given a set of transactions, the prover must show that the hash of a transaction is different from any other hash used to compute $\funcmr$ in the same block. 

An efficient approach to prove this statement is to sort the set of transactions $\{\tx_1, \ldots, \tx_{k'}\}$ (computed in step 1 of the $\mathsf{map}_i$ algorithm described in Figure~\ref{fig:map_reduce_steps}) according to their transaction hash and, then, verify the Merkle path of each transaction following the order of their hash checking. In particular, for every transaction, the prover must show that its transaction hash is strictly greater than the previous transaction in this ordered transactions set.
% This technique ensures that no transaction is processed more than once, as duplicate transactions would generate identical hashes. 

% \stefano{As before, I am confused whether we are retrieving roots for different blocks or for the same block $b_i$}In general, we must modify the relation as follows: 
Hence, in the light of this discussion, we must modify the relation as follows: given an integer $k$ and a set of transaction roots $\{ \roottx^{b_{i_1}}, \ldots, \roottx^{b_{i_m}}\}$ with $1 \le i_1 < i_2 < \dots < i_m \le |B|$\footnoteref{fn:root}, a public claim $x \defeq (\{ \roottx^{b_{i_1}}, \ldots, \roottx^{b_{i_m}}\}, k, r)$ and a private witness $w \defeq (B)$, an oracle server is able to generate a SNARK proof $\pisnark$ when $x$ and $w$ satisfy the following relation $\rel'$:
\begin{itemize}
\item $\funcmr(B) = r$ and,
\item the total number of transaction involved in the computation of $\funcmr(B)$ is $k$ and, 
\item each transaction $\tx$, used to compute $\funcmr$ and contained in a block $b_i$ of $|B|$, is a leaf of the transaction Merkle Tree identified with $\roottx^{b_i}$ and,
\item for each transaction $\tx$, used to compute $\funcmr$ and contained in a block $b_i$ of $|B|$, its hash is different from the hash of any other transactions in $b_i$ used to compute $\funcmr$. 
\end{itemize}

% Formally, let $b_i$ be a block and $T'_{b_i}$ the subset of transactions. Each transaction $\tx \in T'_{b_i}$ has an associated Merkle path $MTP(\tx) = (h_1, h_2, \dots, \mathbb{R}^{b_i}_T)$ corresponding to its position in the transactions Merkle tree of the block. We define the path hash $H(\tx)$ as: $H(\tx) = \mathsf{Hash}(h_1, h_2, \dots, \mathbb{R}^{b_i}_T)$ where $\mathsf{Hash}$ is a collision-resistant hash function. We then define the ordered sequence of path hashes $H(\tx_1) > H(\tx_2) > \dots > H(\tx_M)$, where $M = |T'_{b_i}|$. The circuit enforces this strictly decreasing order by verifying:

% \[
% H(\tx_i) > H(\tx_{i+1}), \quad \forall i \in \{1, \dots, M-1\}
% \]

\paragraph{\textbf{Recursive proving of transaction batches}}
% \item duplicate inclusion - cross batch: the prover submits the same transaction in multiple recursive batches to hide duplicates. ATTENZIONE QUESTO

%The SSLC protocol is designed to remain scalable and efficient, even as the proving claim grows to encompass thousands of transactions within a single sub-task. 

As we said above, $\funcmr$ could be computed over many transactions (in the order of magnitude of thousands or tens of thousands of transactions) and, therefore, computing such proof for such a large claim may require enormous computational resources for the oracle server or even be infeasible.
We want to devise a system that is advantageous for the verifier when the number of transactions on which $\funcmr$ becomes prohibitively large, ensuring, at the same time, that proof generation remains feasible for the prover.  

In particular, the size of the claim scales mostly linearly with the number of transactions included in the computation but remains sublinear to the total number of transactions in a block. 
Specifically, given a block $b_i$ containing $N = |T_{b_i}|$ transactions, we aim to prove a subset of $M < N$ transactions. For each transaction in this subset, the SNARKs prove the correctness of a Merkle proof against a tree of $N$ leaves with root $\roottx^{b_i}$, resulting in $O(M \log N)$ hash computations. % Consequently, memory and computational usage grows superlinearly with the claim.

This is a reasonable scenario because in modern blockchain systems, the number of transactions per block continues to expand to support higher throughput \cite{9862815, 8962150}, generating an SSLC proof for an entire block in a single arithmetic circuit could become infeasibly expensive. To address this challenge, we propose a recursive SNARK composition approach \cite{10.1145/2488608.2488623}. Inspired by the insight of \cite{SP:DVVZ25, EPRINT:DEH24} which demonstrates that generating many smaller proofs is often more efficient than computing a single large proof for an extensive claim, we decide to adopt recursion. This allows the oracle server to generate one \snark per $\funcmr$ while still relaxing the memory requirement by splitting the claim into smaller sub-claims.

In particular, for each $\funcmr$, we partition the set of transactions into batches, one per each block or interval of blocks. For each batch, we generate a \snark which proves that $\mathsf{map}_i$ (or multiple map functions) has been correctly computed on distinct transactions of the block $b_i$, that no transaction has been omitted, and that also verifies the proof of the previous block. Finally, a final \snark proves that $\mathsf{reduce}$ has been correctly computed using as input the outputs computed by the map functions. Such an approach forms a linear recursive chain of proofs, ensuring that all inner proofs are verified if the final proof is valid.

% Formally, given a block $b_i$ and a subset of transactions $T'_{b_i}$, the prover executes a recursive proof system $P(\tau_1, \pi_i^{(\tau_1)}, \tau_2, \pi_i^{(\tau_2)}, ...) \rightarrow \pi_i$ where each inner proof $\pi_i^{(\tau_j)}$ attests to the validity of a batch $j$, and the final proof $\pi_i$ guarantees correctness over the entire set of transactions of block $b_i$.

This recursive approach significantly reduces memory and computational costs, as each proof circuit processes a smaller claim. Furthermore, each recursive step incurs a fixed overhead for verifying the previous proof, which remains independent of the overall transaction set size, making recursion scalable for large proofs.

\subsection{The SSLC Protocol} 
\label{sec:sol-proto}

The SSLC interactive protocol is illustrated in Figures \ref{fig:sslc-q-proto} and \ref{fig:sslc-v-proto}. The protocol involves an SSLC $\lc$ communicating with an Oracle server $\server$ and a set of full nodes $\fn_1, ..., \fn_n$ under the assumption of \textit{existential honesty}. At a high level, the client submits a request for the execution of a map-reduce blockchain query $\funcmr$. The oracle server processes the query using the map-reduce paradigm and returns the computed result along with a proof certifying the correctness of the computation.

To validate the proof, the SSLC interacts with the full nodes to retrieve the information $\st$ necessary for verification, such as the transaction roots of the blocks specified by the prover and the corresponding count of transactions involved in the computation. Finally, the client verifies the proof against the acquired data; if the verification succeeds, the result is accepted; otherwise, it is rejected.

The SSLC protocol is formally defined as the tuple $(\mathsf{Q}(\cdot), \mathsf{Vrfy}(\cdot))$, where:
- $\mathsf{Q}(\cdot)$ represents the \textit{query phase}, in which the SSLC sends the request of computing a function $\funcmr$ to the oracle server and collects the results $r$ along with a SNARK proof $\pi$ and a view of the blockchain $\st$.
- $\mathsf{Vrfy}(\cdot)$ represents the \textit{verify phase}, in which the SSLC connects to $n$ distinct full nodes to collect $\st$ and validate the proof.

\medskip
\noindent\textbf{Query phase.} The protocol $\mathsf{Q}(\cdot)$, described in Figure \ref{fig:sslc-q-proto}, is executed by a light client $\lc$ and an Oracle server $\server$. In the first step, $\lc$ connects to $\server$, sending the description of a map-reduce query $\funcmr$ and the $\server$ with input the blockchain $B$. 

Upon receiving the request, the server executes $\funcmr(B)$ according to the description of Figure~\ref{fig:map_reduce_steps}. %The first step consists of filtering the set of canonical blocks involved in the computation, starting from genesis. These blocks typically include all transactions relevant to the application $\app$, formally defined as the set of transactions satisfying the predicate $\rho(\tx, \app) = 1$. Once the filtering process is complete, the server proceeds with the map-reduce computation by triggering a map-reduce task. Specifically, for each block $b_i$, the server filters the subset of transactions $T'_{b_i}$ and executes $\mathsf{MapReduceTask}(b_i, T'_{b_i})$. These tasks can be processed either sequentially or in parallel, as each sub-task operates independently of the others. The computation of a $\mathsf{MapReduceTask}$ follows the rules outlined in Section~\ref{s:mr-comp}. 

The oracle server also produces a \snark proof certifying the correctness of the computation of $\funcmr$. Specifically, the proof attests that the server correctly applied the function $\funcmr$ over a subset of distinct transactions belonging to a set of transaction roots belonging to a not necessarily subsequent set of blocks (i.e., $\roottx^{b_{i_1}}, \ldots, \roottx^{b_{i_m}}$ with $1 \le i_1 < i_2 < \dots < i_m \le |B|$). 
Once the server ends the computation, the server sends to the SSLC:
\begin{itemize}
    \item The final computed result $r$.
    \item A proof $\pi$ containing:
    \begin{itemize}
        \item A \snark proof $\pisnark$.
        \item A view of the blockchain $\st$, including:
        \begin{itemize}
            \item $\roottx^{b_{i_1}}, \ldots, \roottx^{b_{i_m}}$: the transaction roots belonging to a not necessarily subsequent set of blocks and including the transaction on which $\funcmr$ has been computed;
            \item $k$: the number of transactions processed by $\funcmr$% in block $b_i$.

        %\ivan{intendiamo $k_i$? i numeri sono diversi per ogni blocco no?}
        \end{itemize}
    \end{itemize}
\end{itemize}

\begin{minipage}{\textwidth} 
	\begin{mdframed}[]  % Creates a boxed environment
		\centering \textbf{The SSLC Protocol} \\[0.5em] % Title of the protocol
		%% Query Phase Protocol %%
		\begin{tabular}{@{}p{0.98\linewidth}@{}}  % Table for structured alignment
			\hline\hline
			\multicolumn{1}{c}{\textbf{Query Phase}: $\mathsf{Q}\langle \lc(\funcmr), \server(B) \rangle \rightarrow (r, \pi \defeq (\st \defeq (\{\roottx^{b_{i_1}}, \ldots, \roottx^{b_{i_m}}\}, k), \pisnark))$ \vspace{3pt}} \\
			
			\textbullet\ Protocol steps for the SSLC $\lc$: \\[2pt]
			\hspace{15pt} \textbf{(1) Request.} Connect to the oracle server $\server$: \\
			\hspace{40pt} - send a map-reduce query $\funcmr$\\[2pt]
			\hspace{15pt} \textbf{(2) Update.} Upon receiving $(r, \pi)$ from $\server$: \\
			\hspace{40pt} - store and outputs $(r, \pi)$. \\[4pt]
			
			\textbullet\ Protocol steps for the oracle server $\server$: \\[2pt]
			\hspace{15pt} \textbf{(1) Execute.} Upon receiving a request for $\funcmr$: \\
			\hspace{40pt} - Compute $r \leftarrow \funcmr(B)$ according to Figure~\ref{fig:map_reduce_steps}\\
			\hspace{40pt} - Generate a \snark for the relation $\rel'$ described in Sec.~\ref{sec:solo-proof}\\
			\hspace{15pt} \textbf{(2) Respond.} Send $ (r, \pi \defeq (st \defeq (\{\roottx^{b_{i_1}}, \ldots, \roottx^{b_{i_m}}\}, k), \pisnark))$ to $\lc$\\[4pt]
		\end{tabular}
	\end{mdframed}
\end{minipage}

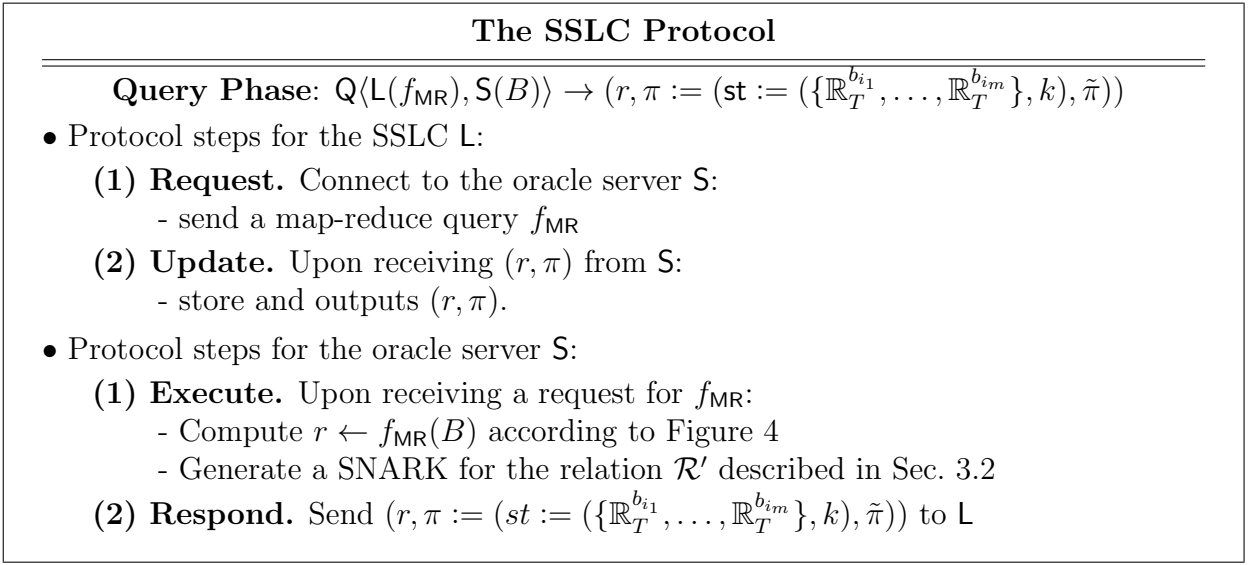
\captionof{figure}{Description of the Query phase of the SSLC protocol.}
\label{fig:sslc-q-proto}

\setlength{\parindent}{15pt} % Restore indentation after mdframed

\medskip

\noindent\textbf{Verify phase.} In the verify phase, the SSLC client $\lc$ interacts with a set of $n$ full nodes $\fn_1, ..., \fn_n$ running the $\mathsf{Vrfy}(\cdot)$ protocol described in Figure \ref{fig:sslc-v-proto}. The main goal of this protocol is to accept or reject the final result provided by the server $\server$ in the query phase.

% A correct map-reduce query must satisfy two properties: \textit{aggregation consistency} for the partial results and \textit{computation completeness} over the set of computed transactions. Additionally, the SNARK proofs attesting to the map-reduce computing tasks must also prove \textit{transaction existence}.

The light client retrieves from the full nodes:  
(i) the total number $k'$ of transactions sent or received by an account $\mathsf{acc}_1$, and  
(ii) the transaction roots $\{\rootfntx^{b_{i_1}}, \ldots, \rootfntx^{b_{i_m}}\}$ from canonical blocks in $B$.  

Full nodes respond if and only if there exists a simple query function $f(\cdot)$ that, given a chain $B$ and additional information (e.g., the blockchain account address $\mathsf{acc}_1$), returns the expected result with a minimal computation task. 
The light client proceeds only if there is no contradiction among the answers of the full nodes. If only one answer differs from on of another full node, the SSLC aborts the protocol. 

Upon receiving the responses from all $n$ full nodes, the client proceeds with the verification steps. 
The first step is to verify the correctness of the \snark proof on the claim $x \defeq (\{ \roottx^{b_{i_1}}, \ldots, \roottx^{b_{i_m}}\}, k, r)$. If the \snark verification succeeds, the SSLC proceeds to the next step. 
%After verifying $\pisnark$, the client checks the total accounting of all transactions used in the computation. Specifically, it verifies that the number of transactions processed across the map-reduce query $\funcmr$ matches the expected total retrieved from full nodes. %Formally, this requires checking whether $\sum_{(b_i, n_i) \in \mathcal{C}} n_i = \mathsf{getTCount}(\mathcal{N})$.
%Ensuring this equality guarantees that no transactions were omitted or falsely introduced in the computation ensuring also the correct computation of $\funcmr$.

\medskip

% As a final step, the light client checks \textit{aggregation consistency} by validating that the aggregation function $\mathcal{A}$ correctly reconstructs the final result $r$ from the set of partial results in $\Gamma$. If the aggregation succeeds, the response is accepted; otherwise, it is rejected.

\begin{minipage}{\textwidth}
\begin{mdframed}[]  % Creates a boxed environment
\centering \textbf{The SSLC Protocol} \\[0.5em] % Title of the protocol

%% Verify phase protocol %%
\begin{tabular}{@{}p{0.98\linewidth}@{}}  % Table for structured alignment
\hline\hline
\multicolumn{1}{c}{\textbf{Verify Phase}: $\mathsf{Vrfy}\langle \lc(\funcmr, r, \pi),  \fn_{1, ...,n}(B) \rangle  \rightarrow b$ \vspace{3pt}} \\

\textbullet\ Protocol steps for the SSLC $\lc$: \\[2pt]
\hspace{15pt} \textbf{(1) Request.} For each $\fn_i$:\\[2pt]
\hspace{40pt} - request transactions count $k$ for account $\mathsf{acc}_1$ on $B$;\\
\hspace{40pt} - request transactions roots for blocks $b_{i_1}, \ldots, b_{i_m}$ from $B$;\\[2pt]
\hspace{15pt} \textbf{(2) Update.} Upon receiving an answer from $\fn_i$: \\
\hspace{40pt} - if $k'$ received by $\fn_i$ is different from a previously received one or \\ \hspace{40pt}  it has received $\bot$, \textbf{abort}; otherwise \textbf{continue}; \\
\hspace{40pt} - if the set $\{\rootfntx^{b_{i_1}}, \ldots, \rootfntx^{b_{i_m}}\}$ received by $\fn_i$ is different from a \\ \hspace{40pt} previously received one or it has received $\bot$, \textbf{abort}; otherwise \textbf{continue};\\[2pt]
\hspace{15pt} \textbf{(4) Verify.} Upon receiving $n$ responses from full nodes:\\
\hspace{40pt} - check if $k' \not = k$; if yes, set $b \rightarrow 0$ and \textbf{return};\\
\hspace{40pt} - check if $\{\rootfntx^{b_{i_1}}, \ldots, \rootfntx^{b_{i_m}}\} \not = \{\roottx^{b_{i_1}}, \ldots, \roottx^{b_{i_m}}\}$; if yes, set $b \rightarrow 0$ and \textbf{return};\\
\hspace{40pt} - check if the $\pisnark$ on the claim $x = (\{ \roottx^{b_{i_1}}, \ldots, \roottx^{b_{i_m}}\}, k, r)$ is correctly verified;  \\ \hspace{40pt}  if no, set $b \rightarrow 0$ and \textbf{return}; \\
\hspace{40pt} - set $b \rightarrow 1$ and \textbf{return}; \\[4pt]

\textbullet\ Protocol steps for the full node $\fn_i$: \\[2pt]
\hspace{15pt} \textbf{(1) Respond Count.} Upon receiving a request for $\mathsf{acc}_1$ on chain $B$:\\
\hspace{40pt} - \textbf{if} $\mathsf{acc}_1$ exists, send the number of transactions sent or received by $\mathsf{acc}_1$ \\ \hspace{40pt} to $\lc$; otherwise, send  $\bot$; \\
\hspace{15pt} \textbf{(2) Respond Block.} Upon receiving a request for a transaction root of $b_i$:\\
\hspace{40pt} - \textbf{if} exists $b_i$ exists, send the transaction root $\rootfntx^{b_i}$ to $\lc$; otherwise, send $\bot$ \\
\end{tabular}
\end{mdframed}
\end{minipage}
\captionof{figure}{Description of the Verify phase of the SSLC protocol.}
\label{fig:sslc-v-proto}

\setlength{\parindent}{15pt} % Restore indentation after mdframed

\paragraph{\textbf{Security and efficiency analysis of the SSLC protocol}} 
% un attaccante può barare facendo: 
% 1) usa uno st' diverso da st ->  corrompere i full node nel dare la risposta che vuole ma per assunzione non riesce
% 2) usando st = st', include alla risposta alla query una transazione che non è nella blockchain mostrando apertura valida -> è in grado di trovare collisione ad una collision resistant hash function
% 3) calcola non correttamente fmr perchè
% a) omette delle transazioni -> avendo numero di txn coivolente e recuperandolo da FN, per existensial honesty, non riesce
% b) utilizza più volte delle txn -> trova collisioni di collision-resistant hash function e per Knowledge Soundness di snark 
% c) devia da fmr -> sicurezza per knowledge soundness di snark

We want now to evaluate the security of the SSLC protocol with respect to the secure querying property described in Section~\ref{subsec:properties}. This property says that it is computationally hard for a malicious oracle server to produce an accepting proof $\pi$ if: 1) the attacker computes  $\funcmr$ on a blockchain view $\st'$ different from the one the SSLC can retrieve by the full nodes or 2) the attacker computes $\funcmr$ on transactions that are not in the blockchain or 3) the attacker does not correctly compute $\funcmr$.

In 1), the attacker uses the following strategy: she generates an accepting \snark proof $\pisnark$ for an ad-hoc realized $\st'$ that differs from $\st$. To achieve this goal, since during $\verify(\cdot)$ $\lc$ contacts $n$ independent full nodes, the malicious server should be able to corrupt all these full nodes. However, if she achieves in doing so, the existential honesty assumption of our system would be violated. Hence, the attacker cannot use this strategy.

In 2), the attacker uses the following strategy: she generates an accepting \snark proof $\pisnark$ for an ad-hoc realized set of transactions. This means that the attacker is able to find at least another transaction such that its hash corresponds to one of the transactions involved in the computation of $\funcmr$. However, if she succeeds in doing so, it means that the attacker can find a collision for a collision-resistant hash function, which is something that happens only with negligible probability. Hence, the attacker cannot use this strategy.

In 3), the attacker deviates from the algorithm to compute $\funcmr$ in the following ways: a) she omits some transactions that should be involved during the computation of $\funcmr$; b) she uses a transaction involved in the computation of $\funcmr$ multiple times; c) she deviates from $f_\mathsf{M}$ or $f_\mathsf{R}$.
Since $\lc$, during $\verify(\cdot)$ protocol, obtains the number $k$ of involved transactions from the full nodes, if the attacker succeeds in a), it means either that she corrupts the full nodes, violating the existential honesty assumption of our system or it can break the knowledge soundness property of the \snark but this happens with negligible probability. 
If the attacker succeeds in b) and c), it means that %either there is a collision for two transaction hashes in the blockchain but this happens with negligible probability or it means that 
the attacker is able to break the knowledge soundness property of the \snark but this happens with negligible probability. Hence, the attacker cannot use this strategy.

Such arguments discussed so far guarantee that our SSLC protocol ensures secure querying.

We are left with showing that our SSLC protocol respects the efficiency property described in Section~\ref{subsec:properties}. 
During $\Q(\cdot)$, the SSLC only needs to send the description of $\funcmr$ to $\server$, therefore, respecting efficient storage, communication, and client computation.
During $\verify(\cdot)$, the SSLC needs to download (from the full nodes) the claim of $\rel'$ on which the \snark is computed which is $(\{ \roottx^{b_{i_1}}, \ldots, \roottx^{b_{i_m}}\}, k, r)$. Then, it reads such a claim while verifying \snark. 
By definition of \snark, the dimension of such proof is sublinear to the size of the transaction list $|T|$, which is included in the witness $B$ of $\rel'$. The reply $r$ to $\funcmr$ is independent of the size of the blockchain because it is aggregated data and $\st$ is composed of a set of transaction roots of the blocks containing the transactions on which $\funcmr$ is computed. By definition of the Merkle Tree, the size of such a set of transaction roots is sublinear to the set of transactions involved in $\funcmr$. Hence, since $\pisnark$, $\st$, and $r$ are sublinear to $|T|$, the efficient storage and communication properties are respected.

Finally, we need to evaluate whether the computational complexity of the SSLC protocol is sublinear to the size of the transaction list $|T|$. Let us suppose that $\funcmr$ is computed over all the transactions in $|T|$. In the SSLC, we compute $\Q(\cdot)$ in an amount of time that is logarithmic with respect to $|T|$ because the SSLC only needs to output the transaction roots of the blockchain. In $\verify(\cdot)$, the SSLC executes the algorithm in logarithmic time with respect to $|T|$ because, in the protocol, it downloads a constant number (i.e., $k$) and the transaction roots, whose number is logarithmic w.r.t. the number of transactions, and it verifies a \snark that by definition is verified in time linear to its claim (i.e., the transaction roots) plus an amount of time sublinear w.r.t. its witness (i.e., $B$  that contains $T$).
\section{Use Case Analysis of SSLC Protocol}
\label{sec:inst}
%\ivan{i nomi delle section in alcuni casi hanno l'iniziale di ogni parola in maiuscolo, uniformare anche secondo lo stile di BCRA}
%\stefano{Maybe is late but --> I expected a comparison with a "naive" approach in which the light client downloads the full transactions from the RPC nodes to compute the complex query. In particular, I am not sure the existing light client solutions that we cite here provide answers to complex queries -- I think they still need a server for that.}

In this section, we show how to adopt the SSLC protocol in two concrete application use cases: a) a Bitcoin SSLC that queries the oracle server to process a statistical analysis on the exchange volume of a given Bitcoin account and b) an Ethereum SSLC to compute statistics over a given time interval on the votes of an on-chain voting system implemented by a Smart Contract. 

To evaluate the performance of our SSLC protocol, we compare it with approaches relying on the Original Nakamoto Light Client (ONLC) and the classes of sublinear light clients (SLCs), introduced in Section~\ref{sec:introduction} which are an optimization of ONLC allowing us to verify the presence of a transaction in a block in sublinear space.

%+The goal of this section is to estimate the number of bytes that a light client must download from full nodes under three scenarios: the Nakamoto traditional light client that typically requires the client to retrieve all block headers; a light client that only stores a sublinear amount of information to keep track of the blocks in the ledger according to the investigations of the state of the art (see Section~\ref{sec:introduction}); the SSLC protocol which leverages \snark succinct proofs and the protocol described in Figure~\ref{fig:sslc-arch} to condense the necessary information into a much smaller data volume, reducing the download size. 

The results of the analysis, summarized in Table~\ref{demo-table}, confirm that our SSLC approach provides a substantial reduction in data transfer, not only minimizing the bandwidth requirements but also enhancing the efficiency and scalability of blockchain clients, especially on resource-constrained devices.

\begin{table}[!h]
\begin{center}
\caption{Comparison of the three approaches on the Bitcoin and Ethereum use cases.}
\label{demo-table}
\begin{tabular}{||c c c c||} 
 \hline
  & ONLC & SLC & SSLC \\ [0.5ex] 
 \hline\hline
 Bitcoin & 726MB & 201MB & 15MB \\ 
 \hline
 Ethereum & 1.1GB & 1GB & 9MB \\
 [1ex] 
 \hline
\end{tabular}
\end{center}
\end{table}

\subsection{The case of Bitcoin} \label{sec:inst-btc}
%In this subsection, we present a  use case showing the potential benefits of our design. 
We first evaluate our solution for monitoring BTC exchange volumes for a given account. Specifically, the oracle server answers queries on the average amount of BTC transferred to or received by a specific account.

To verify the correctness of a response to a query, the application light client executes the following steps:
\begin{enumerate*}[label=\arabic*)]
    \item it retrieves from a blockchain explorer the number of transactions involving a certain address;
    \item it obtains from the server the answer to the query along with a proof showing that these transactions belong to specific blocks (identified with a hash) and how they contribute to the query result;
    \item it queries full nodes to obtain block headers or block hashes for which the proof from the server exists;
    \item it verifies the proof using these trusted hashes;
\end{enumerate*}

In our analysis, we use the Bitcoin address of Satoshi Nakamoto as an example. Three blockchain explorers, namely \href{https://www.blockchain.com/explorer/addresses/btc/1A1zP1eP5QGefi2DMPTfTL5SLmv7DivfNa}{blockchain.com}, \href{https://mempool.space/address/1A1zP1eP5QGefi2DMPTfTL5SLmv7DivfNa}{mempool.space}, and  \href{https://bitaps.com/1A1zP1eP5QGefi2DMPTfTL5SLmv7DivfNa}{bitaps.com}, agree that at block number 881358, Satoshi's address has been involved in 42381 transactions that can be stored in 32 bytes.

In the next, we use block 872028 with 3686 transactions to exemplify the computations necessary to compare the different solutions.  

The oracle response includes the requested average amount (32 bytes) and the proof whose size depends on the circuit and typically ranges between 100 and 150 KB.%; we used Plonky2 which does not require a trusted set-up, as specified in Section~\ref{sec:evaluation}

The RPC calls to answer this complex query are: 
\begin{itemize}
    \item \href{https://www.quicknode.com/docs/bitcoin/getblockhash}{\textbf{\texttt{getblockhash}}}: 101 bytes to retrieve the block hash.
    \item \href{https://www.quicknode.com/docs/bitcoin/getblockheader}{\textbf{\texttt{getblockheader}}}: 613 bytes to retrieve block headers.
    \item \href{https://www.quicknode.com/docs/bitcoin/gettxoutproof}{\textbf{\texttt{gettxoutproof}}}: $\sim$1 KB for the Merkle tree proof of a known transaction ID.
    \item \href{https://www.quicknode.com/docs/bitcoin/getblock}{\textbf{\texttt{getblock}}}: $\sim$10 MB to retrieve the full block.
    \item \href{https://www.quicknode.com/docs/bitcoin/getrawtransaction}{\textbf{\texttt{getrawtransaction}}}: $\sim$3.8 KB for a transaction involving Satoshi.
\end{itemize}

The ONLC must store all block headers; at the time of writing they are 881358 ($\sim534$ MB). Additionally, the client must download: all the 42381 transactions of the examined account ($\sim152$ MB) and all Merkle Tree proofs ($\sim40$ MB).
In total, this approach requires approximately \textit{726 MB}.

\medskip

SLCs approaches need sublinear space. For the sake of simplicity, we assume they use $\log(\text{size of the block header}) = \log(534) = 9.06$ MB. The rest of the operations are the same as in the ONLC. Hence, in total, SLCs require approximately \textit{201 MB}.

\medskip

SSLC significantly reduces storage and computational requirements because
\begin{enumerate*}[label=\arabic*)]
    \item the light client only downloads 42381 block hashes ($\sim$4 MB),
    \item the number of transactions involving the target account (32 bytes) and 
    \item the Plonky2 proof from the oracle server (150 KB).
\end{enumerate*}
Thus, the total downloaded data required to verify the oracle server's answer is \textit{$<$5 MB}. Since we assume that a light client retrieves information from multiple full nodes via RPC, it will download this data multiple times. If three full nodes are contacted, the total download size is approximately \textit{15 MB}.

\subsection{The case of Ethereum} \label{sec:inst-eth}

In this subsection, we present the application of our framework to the verification of statistics on an on-chain voting system governed by a smart contract. %In particular, we consider the case of an oracle server offering the possibility to a SSLC to determine statistics about the participation in the on-chain vote while ensuring correctness through succinct cryptographic proofs.

The smart contract records votes for different candidates as a key(candidate)-value(number of votes) store. In~\ref{apx:sc-voting}, we provide the pseudo-code of a simplified Solidity smart contract for this use case.

Our SSLC aims to verify the average voting participation over a given time interval.

To verify the correctness of the voting participation query, the SSLC executes the following steps:
\begin{enumerate*}[label=\arabic*)]
    \item The SSLC queries the smart contract through an RPC to retrieve the number of voting transactions associated with a candidate.
    \item The SSLC obtains from the oracle server the computed answer along with a cryptographic proof that demonstrates how the retrieved transactions contribute to the final result. The proof guarantees that the transactions and their receipts are included in $m$ blocks.
    \item The SSLC fetches the relevant block headers from a set of full nodes. %using the \texttt{getBlockByNumber} or \texttt{ots\_getBlockDetails} RPC calls.
    \item Given the trusted headers, the SSLC verifies the proof provided by the oracle.
\end{enumerate*}

%We analyze the bandwidth and computational costs associated with our approach. 
We consider an election lasting one day, spanning approximately 7000 blocks, with a candidate receiving 70000 votes, namely an average of 10 votes for every block.

The SSLC fetches the number of votes for a candidate from the smart contract, represented as a 32-byte value. 

The oracle response includes the requested average votes (32 bytes) and the proof, whose
size depends on the circuit and typically ranges between 100 and 150 KB.

The RPC calls to answer this complex query are:

\begin{itemize}
    \item \href{https://www.quicknode.com/docs/ethereum/ots_getBlockDetails}{\texttt{ots\_getBlockDetails}}: $\sim3$ KB to return the details of the block with the specified block number. It is similar to the \texttt{eth\_getBlockByNumber/eth\_getBlockByHash} method, but an optimized version.
    \item \href{https://www.quicknode.com/docs/ethereum/eth_getProof}{\texttt{getProof}}: $\sim7.8$ KB to obtain the account and storage values of the specified account including the Merkle proof.
\end{itemize}

The ONLC needs $\sim21$ MB to store the block headers corresponding to the election period. Additionally, considering the transaction with hash \texttt{0x12e86e5fa174562c61efc03369573ff7058c70229450fd1660e5db4122c4071d} as the reference to compute the following values, it must download: a) all the $\sim70000$ involved transactions, using the \href{https://www.quicknode.com/docs/ethereum/eth_getTransactionByHash}{getTransactionByHash} RPC call ($\sim59$ MB)%\footnote{each transaction requires $\sim 847$ bytes}
b) all the Merkle Tree proofs using the \href{https://docs.nethermind.io/next/interacting/json-rpc-ns/proof/#proof_gettransactionbyhash}{proof\_getTransactionByHash} ($\sim476 MB$) %(assuming that each is 6.8 KB)
%\footnoteref{ftn:txn}; 
, c) all the receipts of the transactions involved using the \href{https://www.quicknode.com/docs/ethereum/eth_getTransactionReceipt}{getTransactionReceipt} ($\sim112 MB$)  %(for each transaction requires $1.6$ KB)\footnoteref{ftn:txn}; 
, d) the related MT proofs (again 476 MB) %\footnoteref{ftn:txn}; 
and finally, e) the proof that a smart contract is in the correct state %a specific state to retrieve the number of transactions she needs to check 
($\sim7.8$ KB). In total, this approach requires approximately \textit{1.1 GB}.

\medskip

SLC in this case downloads $\log(21) = 4.3$ MB. The rest of the operations are the same as in the ONLC. Hence, in total, SLCs require approximately \textit{1 GB}.

\medskip

SSLC significantly reduces storage and computational requirements.  Indeed  it downloads only 1000 block headers (3 MB) and verifies the proof received from the oracle. Even when contacting multiple full nodes (e.g., 3 nodes), the total download remains around $9$ MB. The final verification step of the proof, assuming a Plonky2 proof of 150 KB, results in an overall memory requirement of approximately $9$ MB, a considerable improvement over the other approaches.

\section{Experiments}\label{sec:evaluation}
In this section, we present our experimental evaluation of the SSLC proving system, demonstrating that an oracle server with no resource constraints can efficiently generate \snark proofs attesting to the correctness of map-reduce query executions. Our experiments were conducted on a high-performance server equipped with an Intel(R) Xeon(R) Silver 4216 CPU @ 2.10GHz (64 cores) and 512 GB of RAM. 

%We evaluate two scenarios: (i) an ``optimal case", where the workload is evenly distributed across sub-tasks, and (ii) a ``worst-case", where each map-reduce query processes an exceptionally large set of transactions. 

% Our results indicate that the SSLC remains highly efficient, even when the workload is large, keeping the amount of data retrieved from full nodes limited to only a few pieces of information.

We implemented a \snark circuit in Plonky2, with the source code publicly available on GitHub\footnote{\url{https://github.com/deanstef/awesome-plonky2}}. The implementation simulates the load for the prover and the verifier in realistic scenarios, such as the one of Bitcoin average use case introduced in Section \ref{sec:inst-btc}. Specifically, our implementation verifies Merkle membership proofs while computing the average value of a set of Bitcoin transfer transactions.

Our results show that, even when handling many transactions per circuit, our proof system achieves reasonable memory consumption and proof generation time, while verification remains highly efficient, requiring minimal computational and bandwidth resources. 
% We compare our findings with the empirical evaluations from Section~\ref{sec:inst}, demonstrating that under our assumption, the SSLC remains scalable and efficient, supporting map-reduce queries over hundreds of thousands of transactions.

\subsection{Technical Choices}
\paragraph{Plonky2} We construct our circuit using Plonky2~\cite{plonky2}, a framework for recursive SNARKs. Plonky2 achieves scalability by combining PLONK-based arithmetization~\cite{EPRINT:GabWilCio19} with FRI-based polynomial commitment schemes~\cite{ICALP:BBHR18}. Unlike traditional commitment schemes, FRI provides a transparent, trustless, and hash-based approach, eliminating the need for a trusted setup while enabling efficient polynomial proximity testing. This results in fast proof generation with efficient verification costs. Additionally, Plonky2 leverages the 64-bit Goldilocks field, which is optimized for arithmetic operations on modern hardware. % Unlike Groth16~\cite{EC:Groth16}, which produces constant-size proofs but requires expensive elliptic curve pairings for verification, 
Plonky2 generates polylogarithmic proof sizes while utilizing lightweight hash-based verification. %, making it particularly well-suited for recursive SNARKs.

\paragraph{Circuit Configuration} Our Plonky2 circuit is instantiated using the Plonky2 $\mathsf{CircuitBuilder}$ with a customized configuration to balance performance and overhead. We set $\mathsf{num\_wires = 135}$ to accommodate the circuit’s computational needs without introducing excessive overhead. The transaction Merkle tree is constructed using Plonky2’s $\mathsf{MerkleTree}$ implementation, configured with the Goldilocks field and the Poseidon hash function~\cite{USENIX:GKRRS21}. Each transaction leaf is converted into four field elements to comply with Plonky2’s hashing constraints. Specifically, Poseidon in Plonky2 is optimized for a state width of four elements, ensuring a balance between performance, in-circuit hashing efficiency, and security against collision attacks.

Note that, Poseidon is not yet widely adopted in production-ready blockchains. However, SNARK-friendly hash functions like Poseidon are increasingly seen as essential for blockchain scalability. For example, the transition toward Poseidon has been actively discussed for enabling stateless Ethereum validation via zk-STARKs\footnote{\url{https://vitalik.eth.limo/general/2024/10/23/futures4.html}}.
% ESSENDO UN ARTICOLO NUOVO E NON PUBBLICATO NON LO CITEREI TANTO IL CONCETTO È CHIARO PRIMA and supporting alternative multi-signature schemes~\cite{cryptoeprint:2025/055}. 
Moreover, newer zk-friendly blockchain architectures—including Mina\footnote{\url{https://minaprotocol.com}} and StarkNet\footnote{\url{https://www.starknet.io}}—have already adopted Poseidon as their core cryptographic hash function.

Consequently, we chose not to tailor our implementation to legacy, SNARK-unfriendly hash functions that are likely to become obsolete. Instead, our evaluation is designed to be future-proof, aligned with emerging blockchain developments. That said, it is important to highlight that our SSLC protocol remains compatible with alternative blockchain designs that utilize different hash functions, ensuring flexibility for diverse blockchain ecosystems.

\subsection{Performance Evaluation}
In this subsection, we present the results of our experiments and evaluate the performance and feasibility of our solution. We simulate the load for the prover and the verifier in realistic scenarios (such as the one of Bitcoin average use case) measuring the prover costs associated with generating SNARK proofs for the map-reduce task. 
Additionally, we assess the memory consumed by the SSLC during the verification process. 
Our experiments demonstrate that the solution is practical, enabling clients to verify map-reduce queries involving millions of transactions while downloading only a few megabytes of data and using $\approx$ 2 GB of memory for near-instantaneous verification.

To simulate a heavy workload, we handle a block $b_i$ containing a large transaction set of size $2^{20}$. This means that the transactions will be the hashes of a Merkle Tree with $2^{20}$. The transactions are organized in a Merkle tree using the Plonky2 library $\mathsf{MerkleTree}\text{::}\langle\mathsf{F, PoseidonHash}\rangle\text{::}\mathsf{new}()$, where Poseidon is used as the cryptographic hash function, and $\mathsf{F}$ is instantiated with the Goldilocks field. 

We conduct six experiments, each varying the number of transactions processed in the map-reduce computation. Specifically, we evaluate the test varying the size of the set of transactions on which $\funcmr$ is computed among $10^1, 10^2, 10^3, 10^4, 10^5$ and $10^6$. 
For each experiment, we measure the time required to generate and verify the final proof, the proof size, and the memory consumption for both the prover and the verifier. The experimental results are summarized in Table~\ref{tab:exp-res}.

\begin{table}[h]
    \centering
    \begin{adjustbox}{max width=\textwidth}
    \begin{tabular}{c c | c c | c c} 
        \toprule
        \multirow{2}{*}{\textbf{Workload}} & \multirow{2}{*}{\textbf{Proof Size (KB)}} & 
        \multicolumn{2}{c|}{\textbf{Prover}} & 
        \multicolumn{2}{c}{\textbf{Verifier}} \\ 
        \cmidrule(lr){3-4} \cmidrule(lr){5-6}
        & & Time (s) & Memory (GB) & Time (s) & Memory (GB) \\ 
        \midrule
        $10^1$  & 96.5  & 0.314  & 0.213  & 0.010   & 0.213   \\  
        $10^2$  & 124.8  & 3.418  & 0.320  & 0.020   & 0.270   \\  
        $10^3$  & 156.73  & 29.088  & 1.110  & 0.020   & 0.717   \\  
        $10^4$ & 156.98  & 304.273 & 1.420 & 0.010  & 1.040   \\  
        $10^5$ & 156.98  & 3052.504 & 1.480 & 0.010  & 1.090  \\
        $10^6$ & 156.98  & 30434.426 & 2.060 & 0.010  & 1.660  \\
        \bottomrule
    \end{tabular}
    \end{adjustbox}
    \caption{Performance measurements for different numbers of leaves.}
    \label{tab:exp-res}
\end{table}

In our implementation, we have adopted a recursive proving approach, where at each recursive step our implementation handles a batch of 1000 transactions. 
This batch size simulates a reasonable number of transactions selected by $\funcmr$ per block during the map phase of the algorithm described in Figure~\ref{fig:map_reduce_steps}.
With this configuration, experiments involving up to $1000$ transactions were executed within a single circuit step (i.e., without recursion). For recursive cases, we observed that the average recursion time per step to generate the proof was approximately 30 seconds.

In our Bitcoin average example, each circuit execution sequentially processes the transactions within a block. For each transaction, the circuit performs the following checks: (i) it verifies the Merkle path against the provided root \( \mathbb{R}^{b_i}_T \), (ii) it verifies that $\funcmr$ has selected the relevant transactions (i.e., those sent or received by a specific account), (iii) it verifies that transactions are computed only once (through the ordering technique described in Section~\ref{sec:solo-proof}), and (iv) it computes the average. Additionally, in every recursive step, the circuit verifies the previous proof, which requires loading into the memory a proof for a precedent block along with its corresponding public inputs.

\begin{figure}[h]
    \centering
    % First subfigure: Time Comparison
    \begin{subfigure}[b]{0.49\textwidth}  % Adjust width as needed
        \centering
        \includegraphics[width=\textwidth]{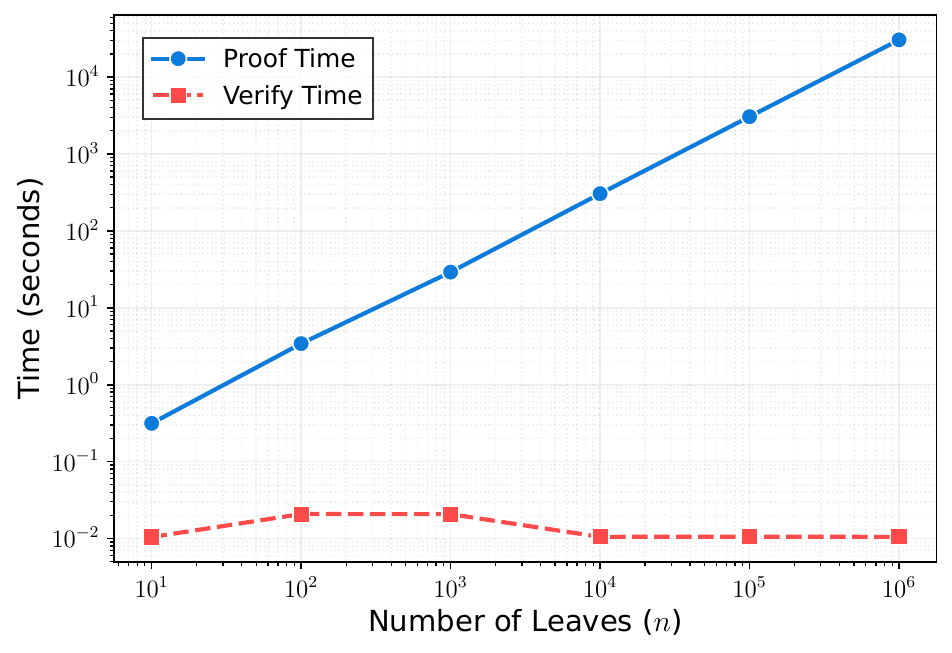} % Replace with actual filename
        \caption{Time Comparison}
        \label{fig:time_comparison}
    \end{subfigure}
    \hfill % Ensures proper spacing
    % Second subfigure: Memory Consumption
    \begin{subfigure}[b]{0.49\textwidth}
        \centering
        \includegraphics[width=\textwidth]{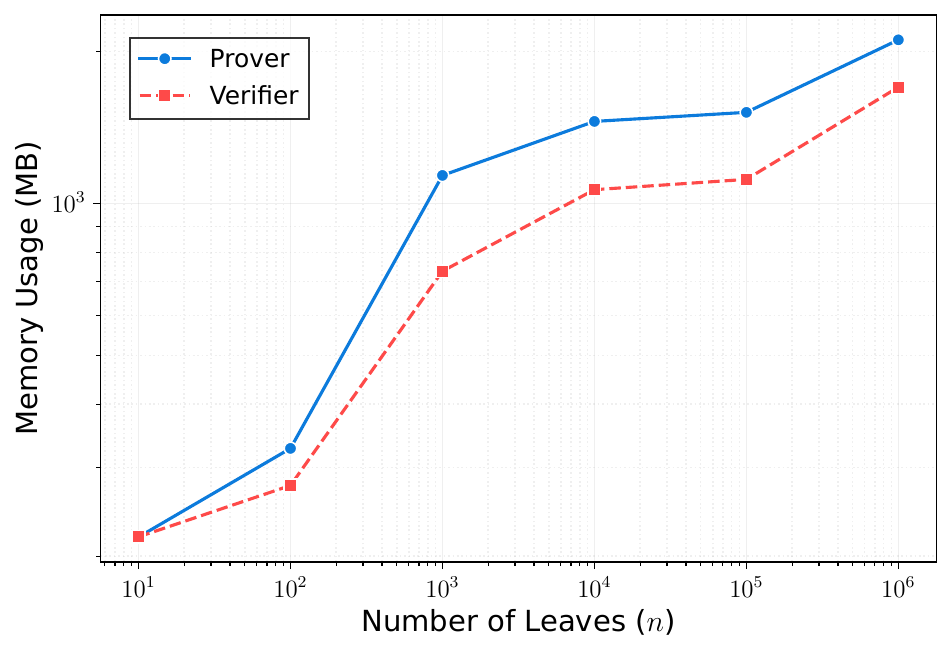} % Replace with actual filename
        \caption{Memory Consumption}
        \label{fig:memory_consumption}
    \end{subfigure}
    \caption{Performance analysis: time and memory comparison.}
    \label{fig:performance_analysis}
\end{figure}

Figure~\ref{fig:time_comparison} presents the measured times for generating and verifying the recursive SNARK. As expected, the prover time increases almost linearly with the number of processed transactions, while the verifier time remains constant, around 10–20 ms. The memory consumption, as illustrated in Figure~\ref{fig:memory_consumption}, follows a similar trend for both the prover and the verifier. % This behavior contrasts with pairing-based proof systems such as Groth16, where verification memory usage remains constant. 
In Plonky2, verifying a recursive SNARK requires storing inner proof elements, including FRI commitments and evaluation points at multiple FRI layers~\cite{ICALP:BBHR18}, resulting in memory overheads similar to those of proving. However, for the SSLC, we observe that even in the worst-case scenario—processing a map-reduce query with $10^6$ transactions—the verifier's memory consumption remains at 1.6 GB, which is within acceptable limits for resource-constrained devices.

The proof size grows up to the batch size. Specifically, we found that a SNARK proof computing $1000$ transactions resulted in a proof size of 156.73 KB. By employing a recursive batching approach, we successfully processed large transaction sets without incurring excessive proof size overhead. Notably, the size of a recursive SNARK proof stabilized at 156.98 KB, remaining constant regardless of the number of recursive steps, with only a minor additional overhead from inner proofs. This result has significant implications for SSLC clients: even in the worst-case scenario, where millions of transactions are processed, the client only needs to download less than 200 KB of data per map-reduce query.

\subsection{Discussion}
%We simulate the cost of a map-reduce query involving $10^6$ transactions. The expected amount of data downloaded by the client from the server would be $X \cdot 156.98$ KB. Since the verifier processes the proofs sequentially, the memory required for verification remains constant at $\approx$ 1.6 GB.

This analysis highlights that for map-reduce queries that involve a large number of transactions distributed across a limited number of blocks, our solution outperforms the ONLC and SLC approaches discussed in Section~\ref{sec:inst}, enabling light clients to verify results with minimal computational and storage resources. 
However, in scenarios where queries span over a large number of blocks, the benefits of this approach become less apparent. In these cases, the SSLC has to download more transaction roots/block hashes, enhancing the amount of communication memory used during the protocol, making it more similar to that of ONLC and SLC. 

% In the following sections, we introduce a technique to distribute proof generation, effectively addressing these limitations. %\stefano{I think something changed from the previous version of the discussion. Now I am not convinced anymore. How do we motivate that for a large X our solution is less effective? And how does the next paragraph help the SSLC? (it helps the prover time for sure, but not the amount of data the SSLC needs to download)}

\paragraph{Prover optimization with parallel SNARKs}
In the current implementation, the prover time scales linearly with the number of $\mathsf{map}_i$ tasks, i.e., it increases by a factor of $X$, where $X$ is the number of blocks involved in the computation. 
Consequently, for applications requiring the processing of a large number of transactions across many blocks, the time required to generate SNARK proofs could become prohibitively long.

However, map-reduce queries exhibit a high degree of parallelism. % In particular, we observe that a valid map-reduce query must ensure aggregation consistency (Section \ref{s:sol-comp}), implying that the partial results are both associative and commutative. This property allows for independent computation of sub-task results, making them well-suited for parallelization. 
Hence, to optimize proving efficiency, we propose leveraging a distributed proving technique, similar to the approach adopted in~\cite{CCS:XZCZZJBS22}. In this scheme, recursive SNARK computations are offloaded to separate hardware units, and the resulting proofs are subsequently aggregated on a central node. This parallelized approach significantly reduces proving time, keeping it close to that of a single $\mathsf{map}_i$ recursive SNARK, irrespective of the number of $\mathsf{map}_i$ executions in parallel.

% \paragraph{Verifier optimization with pairwise aggregated SNARKs}
% SCARICANDO SOLO UNO SNARK PER MAP-REDUCE QUERY ORA NON DOVREBBE SERVIRE QUESTO PARAGRAFO
% In the current design, the verifier downloads and verifies a number of SNARKs equal to the number $X$ of map-reduce sub-tasks. Consequently, as $X$ increases, the light client’s storage requirements also grow. However, the SSLC protocol can be optimized by introducing an additional proving step that aggregates all SNARKs corresponding to each sub-task into a single SNARK.

% A viable implementation involves a pairwise aggregation circuit, forming a binary tree of proofs. This technique, initially proposed in~\cite{EPRINT:DenDu23}, enables the construction of a single aggregated SNARK that verifies multiple inner proofs organized in a tree structure. Specifically, in our case, the aggregation process takes two proofs at a time, verifies their validity, evaluates their respective partial results, and computes an aggregated value. This recursive aggregation continues until a final root proof is generated, attesting to the correctness of the entire map-reduce computation.

% With this approach, the verifier receives a single, constant-size SNARK that encapsulates the correctness of the entire query, eliminating the verification overhead previously introduced by the $X$ individual sub-task proofs.
\section{Acknowledgments}

The work of Marco Zecchini was supported by project SERICS (PE00000014) under the MUR National Recovery and Resilience Plan funded by the European Union - NextGenerationEU. 
The work of Andrea Vitaletti  was partially supported by project  PE11 - MICS (Made in Italy – Circular and Sustainable) under the MUR National Recovery and Resilience Plan funded by the European Union - NextGenerationEU.
Ivan Visconti is member of the Gruppo Nazionale Calcolo Scientifico Istituto Nazionale di Alta Matematica (GNCSINdAM).
The research of Stefano De Angelis and Ivan Visconti is financially supported under the
National Recovery and Resilience Plan (NRRP), Mission 4, Component 2, Investment
1.1, Call for tender No. 104 published on 2.2.2022 by the Italian Ministry of University and Research (MUR), funded by the European Union-NextGenerationEU - Project
Title “PARTHENON”-CUP D53D23008610006 - Grant Assignment Decree No. 959
adopted on June 30, 2023 by the Italian Ministry of Ministry of University and Research
(MUR).

\appendix
\section{Solidity Smart Contract: On-chain Voting Mechanism} \label{apx:sc-voting}
We provide the pseudo-code of a Solidity smart contract that implements the on-chain voting mechanism described in Section \ref{sec:inst-eth}.

\begin{lstlisting}[style=solidity, language=Java, caption={Voting Smart Contract in Solidity}]
contract Voting {
    struct Candidate {
        string name;
        uint voteCount;
    }

    mapping(uint => Candidate) public candidates;
    uint public candidatesCount;

    constructor(string[] memory candidateNames) {
        for (uint i = 0; i < candidateNames.length; i++) {
            candidates[i] = Candidate(candidateNames[i], 0);
            candidatesCount++;
        }
    }

    function vote(uint candidateId) public {
        require(candidateId < candidatesCount, "Invalid candidate");
        candidates[candidateId].voteCount++;
    }

    function getVotes(uint candidateId) public view returns (uint) {
        require(candidateId < candidatesCount, "Invalid candidate");
        return candidates[candidateId].voteCount;
    }
}
\end{lstlisting}

The contract defines a simple voting system where candidates are stored in a mapping, each identified by an index. Each candidate has a name and a vote count. The constructor initializes the contract with a predefined list of candidates, each starting with zero votes. 

To allow users to vote, the contract provides a \texttt{vote} function that takes a candidate ID as input and increments the corresponding vote count. A \texttt{require} statement ensures that the candidate ID is valid before updating the count. Additionally, a \texttt{getVotes} function allows users to retrieve the number of votes a candidate has received. 

This implementation provides a minimal and functional voting mechanism but does not include security features such as preventing double voting, authentication, or encryption of votes. However, we can use it to show the incremental counter per candidate as the crucial state parameter for a SSLC. In particular, the counter represents the number of vote transactions emitted per candidate and can be used by a light client in the SSLC protocol to infer computation completeness of the map-reduce query.

\end{document}